\documentclass[journal,11pt,onecolumn,english,comsoc]{IEEEtran}
\usepackage[utf8x]{inputenc}
\usepackage[T1]{fontenc}    % use 8-bit T1 fonts
\usepackage{hyperref}       % hyperlinks
\usepackage{url}                 % simple URL typesetting
\usepackage{booktabs}       % professional-quality tables
\usepackage{amsfonts}       % blackboard math symbols
\usepackage{nicefrac}       % compact symbols for 1/2, etc.
\usepackage{microtype}      % microtypography
\usepackage{lipsum}
\usepackage{graphicx}
\usepackage[plain]{algorithm}
\usepackage{algpseudocode}
\pdfoutput=1
\begin{document}
	
\title{Energy Efficient Power and Channel Allocation in Underlay Device to Multi Device Communications}
\author{Mariem~Hmila, Manuel~Fernández-Veiga,~\IEEEmembership{Senior~Member,~IEEE}, Miguel~Rodríguez-Pérez,~\IEEEmembership{Senior~Member,~IEEE}, and Sergio Herrería-Alonso
	\IEEEcompsocitemizethanks{
		The authors are with the Department of Telematics Engineering, University of Vigo, Spain (e-mail: \{meriame,mveiga,miguel,sha\}@det.uvigo.es).
	}
	\thanks{This work was supported by the European Regional Development Fund
		(ERDF) and the Galician Regional Government under agreement for
		funding the atlanTTic Research Center for Telecommunication
		Technologies (atlanTTic), and by the ``Ministerio de Economia,
		Industria y Competitividad'' through the project TEC2017-85587-R of
		the ``Programa Estatal de Investigación, Desarrollo e Innovación
		Orientada a los Retos de la Sociedad'' (partly financed with FEDER
		funds).}}

\markboth{Journal November~2018}
{Shell \MakeLowercase{\textit{et al.}}: Journal Paper}
\maketitle
\begin{abstract}
	In this paper, we optimize the energy efficiency
	(bits/s/Hz/J) of device-to-multi-device (D2MD) wireless communications.
	While the device-to-device scenario has been extensively studied to
	improve the spectral efficiency in cellular networks, the use of multicast
	communications opens the possibility of reusing the spectrum resources
	also inside the groups. The optimization problem is formulated as a mixed
	integer non-linear joint optimization for the power control and allocation
	of resource blocks (RBs) to each group. Our model explicitly considers
	resource sharing by letting co-channel transmission over a RB (up to a
	maximum of $r$ transmitters) and/or transmission through $s$ different
	channels in each group. We use an iterative decomposition approach, using
	first matching theory to find a stable even if sub-optimal channel
	allocation, to then optimize the transmission power vectors in each group
	via fractional programming. Additionally, within this framework, both the
	network energy efficiency and the max-min individual energy efficiency are
	investigated. We characterize numerically the energy-efficient capacity
	region, and our results show that the normalized energy efficiency is
	nearly optimal (above 90 percent of the network capacity) for a wide range of
	minimum-rate constraints. This performance is better than that of other
	matching-based techniques previously proposed.
\end{abstract}


\section{Introduction}
\label{sec:intro}
A number of different technological advancements are being
considered for realizing the expected capabilities of 5G
networks~\cite{andrews2014will}, in terms of data volume, throughput and
users' density. These include massive MIMO, mm-wave communications, small
cells, content caching and device-to-device (D2D) communications, among
others. In particular, D2D is a conceptually simple technique that allows
users in close proximity to communicate directly without the intervention of a
central entity like a base station (BS) or an access point (AP). This
short-range communication mode provides high data rate, low latency, high
energy efficiency and increased system capacity. D2D communications can take
place in-band or out-band~\cite{mach2015band}, the first type when a
D2D-enabled device shares the licensed spectrum with cellular users (underlay
communication), the second case if dedicated bandwidth is reserved for D2D
devices, which then would form an overlay network. One of the key performance
metrics for optimizing the design and operations in cellular networks and for
using efficiently the physical resources is the energy efficiency
(EE)~\cite{Zappone16}. Energy efficiency is the normalized data transmission
rate (bits per second per hertz) divided by the amount of energy used for
achieving that reliable transmission rate. The design of
networks for optimizing EE usually requires a shift from the techniques used
for optimizing other types of resources~\cite{di2018system} since network EE
turns out to be a complex function of the network structure. Generally, EE
may depend on (i) the location of transmitters and receivers; (ii) the
resource allocation (shared/non-shared); (iii) the aggregate interference,
which in turn also depends on the power control algorithms; (iv) and the
direct transmission mode among devices (unicast or multicast).

In this paper, we address the system-level optimization and design of the EE
in an underlay D2D network with multicast transmitters (D2MD). Multicast does
not waste transmission opportunities when multiple receivers consume the same
flow of data, thus offering higher potential for energy savings. Similarly,
underlay communications enable the D2MD devices to share the channels
allocated to ordinary cellular users, and to use less spectrum, but with the
drawback that the level of interference is increased in comparison to overlay
communications. Though in underlay D2MD the spectral efficiency can be higher,
the main challenge is how to mitigate interference among users sharing the
same channels or resource blocks (RBs).

\subsection{Related Work}
Several works in the literature have addressed the coexistence of D2D underlay
transmissions in a cellular network. In~\cite{Dinh-Van2015}, up-link resource
allocation and power control algorithms are proposed to enhance system
spectral efficiency. Here, D2D pairs are divided into clusters based on the
distance between them and each cluster uses a single RB. The authors test
several distance-based criteria for the pairing of D2D-enabled devices and
cellular users, in order to keep the transmission power below the maximum
threshold. Their simulation results show that D2D coexistence increases the
total throughput. Joint power and resource allocation were also investigated
in~\cite{Alamouti2014} to maximize the EE of cellular users. The problem was
solved in two stages: the first is the power control problem, then the outcome
is used to create a bipartite graph to model the resource allocation, which is
solved with the well known Hungarian algorithm~\cite{Kung55}. A more evolved
situation appears in~\cite{Zhang2016}, where hyper-graph theory is applied.
Hypergraphs allow forming multiple edges between two vertices---including
self-edges---, and this feature facilitates the tracking and control of
cumulative interference to maximize system sum-throughput subject to power
constraints. The hyper-graph construction is then combined with a greedy
coloring algorithm to determine the final allocation. The results show
significant improvements when compared to the typical approach with simple
graphs, and demonstrate that simple graphs cannot capture all the
relationships among the interfering sources. Other works
as~\cite{Alamouti2014,Zhou14}, propose energy-efficient resource allocation
algorithms subject to diverse performance objectives and transmission power
constraints. For more details we refer to the survey~\cite{Ali2017}.

All these previous works only consider unicast transmission. However, underlay
multicast or device-to-multi device (D2MD) helps to reduce overhead signals
and increases network capacity with fewer resources. Nevertheless, besides
controlling interference, D2MD poses its own challenges such as the selection
of the head cluster, or the strategy for forming the groups. Only a few works
have analyzed EE and interference mitigation in
D2MD. In~\cite{meshgi2015joint}, the problem of joint power control and
resource allocation for D2MD was studied, aiming at the maximization of system
throughput.  A two-stage decomposition is followed, where the first
sub-problem is to identify the optimal power allocation and the feasible
channel subset, and the second stage consists in finding a good allocation of
channels through a bipartite graph and the Hungarian algorithm. This work is
extended in~\cite{meshgi2017optimal}, which considers a scenario where
multiple D2MD groups can use multiple channels. The paper provides a
comparison between a greedy and a heuristic algorithm. With a similar concept,
the authors in~\cite{Li2016} modelled the problem using a multi-objective
optimization framework with weighted factors to minimize energy consumption
while maximizing the number of served links in D2MD
groups. In~\cite{Hamdi2017}, the problem was formulated as a mixed integer
non-linear problem (MINLP) where the D2MD groups can use all the cellular
channels to maximize the minimum throughput. This was also heuristically
solved in two stages, where the second one uses a genetic algorithm. A
different approach is presented in~\cite{Zhou2016}, where
resource allocation is done using matching theory while power control is
solved using fractional programming and the Dinkelbach's algorithm. The main
goal is to maximize the energy efficiency of individual users. Based on
this brief review, we can classify resource allocation in the related work
according to: (i) the number of users per channel; (ii) the amount of RBs a
D2MD group can use. These criteria give four possible types or scenarios:
\begin{enumerate}
	\item Type 1: each D2D pair/group can only use one RB, and vice-versa.
	
	\item Type 2: a D2D pair/group can distribute its rate over distinct RBs,
	but a cellular user (CUE)  shares its resources with a single D2D pair/group.
	
	\item Type 3: a D2D pair/group can use one RB. However, CUE channel
	can be shared among various D2D pairs/groups.
	
	\item Type 4: a RB can be shared among different D2D pairs/groups which can
	use many RBs.
\end{enumerate}
Please, refer to Table~\ref{table:stateOfart} for a summary of the relevant
literature in the context of D2D and D2MD system optimization.

\begin{table*}[t]
	\centering
	\small
	\caption{\label{table:stateOfart} D2D Resource Allocation (RA) and Power Control (PC): State of the Art}
	
	\begin{tabular}{ccllll} \hline
		\textbf{Ref.} & \textbf{Scenario} & \multicolumn{1}{c}{\textbf{Approach}} & \textbf{Model} & \textbf{Problem} & \multicolumn{1}{c}{\textbf{Objective}}  \\ \midrule
		\cite{Dinh-Van2015} & 3 & Optimization & D2D & RA, PC
		& Spectral Efficiency \\ 
		\cite{Alamouti2014} & 1 & Optimization; Graph theory 
		& D2D & RA, PC & CUE Energy Efficiency \\ 
		\cite{Zhang2016} & 2 & Optimization; Graph theory 
		& D2D & RA, PC & Sum Throughput \\ 
		\cite{Hmila2017} & 3 & Optimization; Stochastic geometry 
		& D2MD & PC & Users/System Energy Efficiency \\ 
		\cite{meshgi2015joint} & 1 & Optimization; Graph theory 
		& D2MD & RA, PC & Sum Throughput \\ 
		\cite{meshgi2017optimal}& 4 & Optimization; Heuristic algorithm  & D2MD
		& RA, PC & Sum Throughput \\ 
		\cite{Li2016}  & 3 & Optimization  & D2MD & RA, PC
		& Spectral Efficiency \\ 
		\cite{Hamdi2017} & 3 & Optimization; Genetic algorithm 
		& D2MD & RA, PC & D2D Minimum Throughput \\ 
		\cite{Zhou2016} & 3 & Optimization, Matching
		theory & D2MD  & RA, PC &
		Individual energy efficiency
		\\ \hline
	\end{tabular}
\end{table*}

\subsection{Contributions}
Observing Table~\ref{table:stateOfart}, clearly the main objectives are either
to maximize the sum-throughput or the system capacity, with just a single work
investigating EE for D2MD communications. In our previous
work~\cite{Hmila2017}, we found that, when the aggregate interference is well
controlled, a channel can hold up to a few tens of D2MD groups. In such case,
the sum-throughput continuously increases as the number of users grows,
whereas both the system and the individual EE decrease significantly. In this
paper, we analyze the trade-off between EE and throughput in D2MD
communications, and in this context make the following contributions:
\begin{enumerate}	
	\item We introduce in our formulation the possibility of splitting a
	transmission over different channels and of reusing a channel among
	different transmitters, up to a predetermined maximum. This is equivalent to
	considering the fractional reuse of resources, a feature not considered in
	other related works. Moreover, we prove that the resulting joint
	optimization problem is NP-hard.
	
	\item We propose a two-stage decomposition approach for solving the resource
	allocation and power control sub-problems. Only the first sub-problem is
	solved sub-optimally, using matching-theory concepts. For the power
	allocation, a classical optimization technique is employed, once the
	sub-problem is recast as a fractional programming case. The
	two sub-problems are solved iteratively, not sequentially, as in other
	works (e.g.,~\cite{Zhou2016}). In doing so, we always select the best
	transmission power for a given channel assignment. In addition, this
	comprises in a common framework the maximization of the global network EE
	(GEE) and the maximization of the minimum individual EE, i.e., per device
	(MEE).
	
	\item We present numerical results showing that this mathematical methodology
	allows us to design the network for maximizing EE while preserving rate and
	power constraints. System performance, in terms of EE, is close to optimal
	and the computational complexity of the procedure is reasonable. We compare
	our results with an exhaustive search algorithm to assess its accuracy
	and also with the iterative matching-based algorithm
	from~\cite{Zhou2016}. We obtain better both EE and sum-rates
	than~\cite{Zhou2016}.
\end{enumerate}

The rest of the paper is organized as follows. In
Section~\ref{sec:system-model}, we introduce system model. Problem formulation
is discussed in Section~\ref{sec:Problemformulation}. Later, we present
resource allocation in Section~\ref{sec:ResourceAllocation}, followed by power
control in Section~\ref{sec:PowerControl}. Finally, we discuss simulation
results in Section~\ref{sec:simulations} and conclude in
Section~\ref{sec:conclusion}.
	
\section{System Model} 
\label{sec:system-model}
Our model considers a single cell/single tier network with one central entity
located at the center. Users are randomly distributed over the cell area. On
the up-link (UL), $M$ cellular users (CUEs) transmit on $M$ orthogonal
communication sub-channels or resource blocks. Other works
(e.g.,~\cite{Afshang16}) have argued that the reuse of the down-link channel
in wireless networks has less performance gains than sharing the up-links, and
it requires more complex coordination between the end-user devices. Thus, we
focus on the UL, and assume users will be grouped into $S$ multicast clusters
$\mathcal{S}_k$, $k = 1,...,S$, that can reuse the same communication channels
allocated to the CUEs for direct communication among their members. Each of
these device-to-multi-device (D2MD) groups has only one designated transmitter
and $s_k = |\mathcal{S}_{k}| - 1$ receivers. The special case where
$|\mathcal{S}_{k}| = 2$ reduces to a unicast communication or D2D
pair~\cite{meshgi2017optimal}. In this model, the BS is hampered by the
interference caused by the co-channel D2MD transmitters, and the $s_k$
receivers in group $k$ suffer from interference caused by the cellular users
and other transmitters of D2MD groups sharing the same resource block. At a
given D2MD receiver $r \in \mathcal{S}_k$ which receives on the sub-channel
$m$, the Signal-to-Interference-and-Noise Ratio (SINR) is
	\begin{equation}
	\label{eq:sinr-d2d} 
	\gamma_{k, r}^{(m)} = \frac{h_{k,r}^{(m)} p_{k}^{(m)}}{\sigma^2  + p^{(m)}
		\beta_{k,r}^{(m)} + \sum_{j \neq k} \delta_{j,m} p_{j}^{(m)} h_{j}^{(m)}},
	\end{equation}
	
	where $h_{k,m}^{(r)}$ are the channel coefficients for the link between the
	transmitter-receiver pair $(k, r)$ when using the $m$-th RB;
	$\beta_{k,r}^{(m)}$ is the channel gain for the path between CUE
	transmitter $m$ and receiver $r$ in group $k$, which uses transmission power
	$p_{k}^{(m)}$; $p^{(m)}$ is the transmission power of CUE user $m$;
	and $\delta_{j,m}$ is the indicator variable for group $j$
	using channel $m$.  All the channel coefficients are assumed to be
	independent of the users' transmit powers, and only dependent on the physical
	properties of propagation channels, see e.g~\cite{Calcev07}. At the $m$-th
	CUE, the SINR is similarly expressed as
	\begin{equation}
	\label{eq:sinr-cue}
	\gamma^{(m)}(\mathbf{p}^{(k)}, p^{(m)}) = \frac{h^{(m)} p^{(m)}}{\sigma^2 +
		\sum_k \delta_{k,m} p_{k}^{(m)} h_{k}^{(m)}}, \quad m = 1, \dots, M 
	\end{equation}
	where $h^{(m)}$ is the channel coefficient $m$ to the base station, $p^{(m)}$
	is the transmitted power, $h_{k}^{(m)}$ is the link gain from the transmitter
	in D2D group $k$ to the cellular base station on channel
	$m$. In (1) and (2), $\delta_{k,m}$ are the
	indicator variables, that is, $\delta_{k,m} = 1$ if D2D group $k$ uses channel $m$, $0$
	otherwise, defined for all $k, m$. We assume that all the channels are AWGN
	channels with noise power $\sigma^2 = N_0 W$, where $N_0$ is the noise power
	density and $W$ is the channel bandwidth, and that all the receivers decode
	the received signal treating interference as noise~\cite{ElGamal13}. Under
	these assumptions, the normalized transmission rate in bits per second per Hz
	for CUE $m$ is the ergodic channel capacity~\cite{Shannon48}
	\begin{equation}
	\label{eq:rate-cue}
	r_m = \log_2 (1 + \gamma^{(m)}), \quad m = 1, \dots, M.
	\end{equation}
	In any D2MD group $k$, the multicast transmission rate is constrained by the
	weakest receiver, the one with poorest channel quality. In addition, we
	account explicitly for the aggregated received rate in group $k$ such as
	$k = 1, \dots, K$ that depends on the number $s_k$ of receivers per group. So,
	
	\begin{IEEEeqnarray}{rClrCl}
		\label{eq:rate-d2d}
		R_k & = & \sum_{m = 1}^M \delta_{k,m} s_k \log_2 (1 + \min_{r \in \mathcal{S}_k}
		\gamma_{k,r}^{(m)})\quad & = & \quad \sum_{m = 1}^M \delta_{k,m} s_k \min_{r \in \mathcal{S}_k} \log_2(1 +
		\gamma_{k,r}^{(m)}),
	\end{IEEEeqnarray}
	
	\section{System Optimization}
	\label{sec:Problemformulation}
	
	In this Section, we present the performance metrics used for determining the
	energy efficiency of the system, first introduced
	in~\cite{Zappone16,Zappone16b}. Next, we pose the corresponding optimization
	problems for maximizing the energy efficiency of the wireless network. In
	these problems, we assume that power usage can be modeled as an affine
	function.
	
The energy efficiency (EE) of a given user (in bits/s/Hz per joule) is the
ratio of the achievable normalized transmission rate and the total consumed
energy:
\begin{equation}
\label{eq:ee-cue}
\eta_m \triangleq \frac{r_m}{\tau_{m} + p_m}, \quad m = 1, \dots, M
\end{equation}
for a CUE user; and
\begin{equation}
\label{eq:ee-d2d2}
\zeta_k \triangleq \frac{R_k}{\tau^\prime_{k} +
	\boldsymbol{\delta}_k^T\mathbf{p}^{(k)}_1}, \quad k = 1, \dots, K
\end{equation}
for a D2D user.

In the definition $\tau_{m}$ (respectively, $\tau^\prime_{k}$) is the
transmitter circuit power at rest (i.e., when there is nothing to transmit),
$\mathbf{p}_k = (p_{k}^{(1)}, \dots, p_{k}^{(M)}) \in \mathbb{R}_+^M$ is the
allocated power vector of the head cluster over the $M$ channels,
$\boldsymbol\delta_k =(\delta_{k,1}, \dots, \delta_{k,M})^T$ is the vector of
channel assignments used by transmitter $k$, and ${\cdot}_1$ denotes the
$\ell_1$-norm. We assume that $p^{(m)}$ and $\mathbf{p}_k$ satisfy individual
power constraints ${\mathbf{p}_k}_1 \leq \overline{p}^{(k)}$ for
$k = 1, \dots, K$, and $p^{(m)} \leq P^{(m)}$ for all $m$. We also want to
consider the constraint that there are minimum transmission rates
$\underline{r}_m$ and $\underline{R}_k$ for all the devices, both the CUEs and
the D2MD users,
\begin{equation}
r_m \geq \underline{r}_m, \quad\text{and } R_k \geq \underline{R}_k, \quad
\forall k, m.
\end{equation}
Note that the definition of energy efficiency is given on a per-user basis,
not for the system as a whole. The system's global energy efficiency (GEE)
$\eta$ is simply the ratio between the aggregated rate and the total power
needed. So, if $\mathbf{r}$ and $\mathbf{R}$ are the vectors of rates for CUE
and D2MD groups, respectively, then
\begin{IEEEeqnarray}{rCl}
	\label{eq:network-ee}
	\eta & := & \frac{{\mathbf{r}}_1 +
		{\mathbf{R}}_1}{\tau + \sum_k \boldsymbol{\delta}^T
		\mathbf{p}_k + {\mathbf{p}}_1}.
\end{IEEEeqnarray}
Here, $\tau := \sum_m \tau_{m} + \sum_k \tau^\prime_{k}$ indicates the total
power used by the circuitry of the devices in the network. However, in some
scenarios, it could be useful to focus on the worst-case performance of a
given device, for instance whenever the limited battery of devices could be
particularly stringent, as in wireless sensor networks. Thus, following a
max-min fairness criterion, the (generalized weighted) minimum EE in the
system is

Let
$\boldsymbol\omega = (\omega_1, \dots, \omega_M) \in  \mathbb{R}_{+}^M$
and
$\boldsymbol\theta = (\theta_1, \dots, \theta_K) \in
\mathbb{R}_{+}^K$
be two arbitrary weight vectors. The
$(\boldsymbol\omega, \boldsymbol\theta)$-weighted energy efficiency
(WEE) is
\begin{equation}
\eta_{\mathsf{WEE}} := \min \{ \min_m \omega_m \eta_m, \min_k
\theta_k \zeta_k \}. 
\end{equation}

The uniform choice
${\boldsymbol\omega}_2 = {\boldsymbol\theta}_2 = 1$, giving equal
weight to every user, yields max-min fairness as the optimization
criterion. Now, with the performance metrics already defined, we can formulate
two straightforward energy efficiency optimization problems.

\noindent\textbf{GEE - Global Energy Efficiency}

\begin{equation}
\label{eq:gee}
\max_{\mathbf{p}  \in \mathcal{P}} 
\frac{\mathbf{r}_1 + {\mathbf{R}}_1}{\tau +
	{\mathbf{p}}_1 + \sum_k \boldsymbol\delta_k^T \mathbf{p}_k}
\end{equation}
with $\mathcal{P}$ the feasible set of power vectors. This is the set defined
by the constraints
\begin{IEEEeqnarray}{rCl}
	\sum_{m} p_{k}^{(m)} & \leq & \overline{p}_k,\quad k = 1, \dots, K
	\IEEEyessubnumber* \label{gee:c1} \\
	p^{(m)} & \leq & P^{(m)}, \quad \forall m \in \mathcal{M} \label{gee:c2} \\
	\underline{R}_k & \leq & R_k, \quad \forall k \in \mathcal{K} \label{gee:c3} \\
	\underline{r}_m & \leq & r_m, \quad \forall m \in \mathcal{M} \label{gee:c4} \\
	\sum_{m}\underline{R}_{k,m} & \leq & R_k, \quad \forall k \in
	\mathcal{K} \label{gee:c8} \\
	\boldsymbol\delta_k \in \{0, 1 \}^M, & & {\boldsymbol\delta_k}_1 \leq
	s \quad \forall k \in \mathcal{K} \label{gee:c5} \\
	& & {\boldsymbol\delta_{\cdot,m}}_1 \leq r \quad \forall m \in
	\mathcal{M} \label{gee:c6} \\
	& & (\mathbf{p}, \mathbf{p}_1, \dots, \mathbf{p}_K) \in
	\mathbb{R}_+^{(K + 1) \times M} \label{gee:c7}.
\end{IEEEeqnarray}

Most of these constraints are natural according to our setting.
First,(10a)-(10b)bound the maximum power per user;
constraints (10c)-(10d) enforce the minimum rate conditions,
where the rates have been defined in (3)
and(4) and depend on the particular channel
allocation variables through the interference terms appearing in the SINR;
constraint(10f) introduces the maximum split factor~$s$ for every
D2MD group: it lets a transmitter to use simultaneously up to $s$ RBs,
distributing its power among them to satisfy a given rate
constraint(10e); conversely,(10g) is the maximum reuse
factor $r$ per resource block: this sets a maximum to the number of
simultaneous transmitters that a sub-channel supports;(10h) is the
non-negativity of all the power vectors. Since the optimization problem
combines integer constraints (10f -10g) and real
variables (the coupling constraints 10c-10d, and the
coupling variables $\{ \mathbf{p}_k \}$), we deal with a mixed-integer
non-linear optimization problem (MINLP), which is hard to solve. As a final
remark, observe that (10c -10d) are given per-user. This
implies that the assignment problem embedded into GEE is to find the optimal
assignment of a subset of D2MD clusters to the channels, not only the number
of D2MD clusters which use that channel like in~\cite{Xu12, Xu13}.

\noindent\textbf{Problem MEE - Minimum Energy Efficiency}

\begin{equation}
\label{eq:minimum-ee}
\max_{\mathbf{p} \in \mathcal{P}} \eta_{\mathsf{WEE}}
\end{equation}
subject to the same constraints (10a)-(10h). 

We notice here that the RB sharing and power control problems stated
in (10) and (11) are NP-hard. This fact holds not
only because they belong to the class of MINLP problems, which are generally
NP-hard, but because there exists an explicit reduction of them (in polynomial
time) to a well-know NP-hard problem, namely the integer partitioning
problem. We refer the reader to the Appendix for the proof.  The above
mathematical framework is general enough to include several particularizations
of interest:

\begin{enumerate}
	\item If $s = r = 1$, a D2MD group can only use a single CUE, and a RB
	supports at most one group. Therefore, in this case transmissions are
	orthogonal and do not interfere with each other. The values $R_k$ and $P_k$
	in constraints (10a) and (10c) refer to minimum rate and
	maximum power over the chosen RB, respectively, and (10e) is
	inactive.
	
	\item If $s = 1, r > 1$, multiple (up to $r$) D2MD groups are allowed to share
	and use a single RB. Here, we have inter-group and accumulated interference
	at the CUE and D2MD receivers. Now, the constraints (10a)
	and (10c) bound to minimum rate and maximum power over the chosen
	RB, and (10e) is again inactive.
	
	\item If $r = 1, s > 1$, a set of $s$ resource blocks, at most, can be
	allocated to a D2MD group. So, this scenario is similar to the first one in
	terms of interference, yet constraints (10e),( 10a)
	explicitly consider the distributed rate $R_k$ and the maximum total
	transmission power $P_k$ over the used RBs. In addition, (10c)
	bounds the minimum rate per channel to avoid having extremely low data rates
	on any individual RB. Finally, each time a D2MD uses a RB it consumes a
	specified amount of circuit power thus, $\tau_{m}$ is a vector.
	
	\item The choice $r > 1, s > 1$ is the general case. Here, a D2MD is allowed
	to distribute its transmission power and rate over all the $s$ RBs and a CUE
	user is allowed to share its RB with $r$ D2MD groups, at most. All the
	constraints in the optimization problems might be active.
\end{enumerate}

	\begin{table*}
		\centering
		\small
		\caption{\label{d2d:symbols} Notation and Symbols}
		
		\begin{tabular}{llll}\hline
			\textbf{Symbol} & \textbf{Definition}& \textbf{Symbol} & \textbf{Definition}\\ \midrule
			BS & Base Station & $r_m$ & rate of the $m$-th CUE \\
			CUE & Cellular User Equipment & $R_k$ & rate of the $k$-th group \\
			RB & Resource Block &     $\underline{r}_m$ & minimum target rate for user $m$ \\ \cmidrule{1-2}
			$K$ & number of groups/clusters &     $\underline{R}_k$ & minimum target rate for group $k$ \\
			$M$ & number of cellular users, RBs or channels &     $\overline{p}_k$ & maximum transmission power in group $k$ \\
			$\mathcal{M}$ ($\mathcal{K}$) &
			set of cellular users/channels (clusters) &     $P^{(m)}$ & maximum transmission power in channel $m$  \\ \cmidrule{3-4}
			$\sigma^2$ & noise power  &     $\tau_{m}$ & fixed circuit power, user $m$ \\
			$s$ & split factor: max. number of simultaneous RBs for a transmitter & $\tau^\prime_{k}$ & fixed circuit power, transmitter in group $k$ \\
			$r$ & reuse factor: max. number of transmitters in a channel/RB  &     $\eta_{\mathsf{GEE}}$ & global energy efficiency \\ \cmidrule{1-2}
			$\gamma_{k,r}^{(m)}$ & SINR of D2MD receiver $r$ in group $k$ on channel $m$ &     $\eta_{\mathsf{WEE}}$ & (weighted) minimum energy efficiency \\\cmidrule{3-4}
			$\gamma^{(m)}$ & SINR at CUE transmitter $m$ &     $h^{(m)}$ & link gain from transmitter $m$ to BS \\
			$h_{k,r}^{(m)}$ & channel gain between tx. $k$ and receiver $r$ on channel $m$ &    $p_{k}^{(m)}$ & transmission power of $k$ on channel $m$ \\
			$\beta_{k,r}^{(m)}$ & channel gain from CUE transmitter $m$ to receiver $r$ on
			group $k$ &    $p_m$ & transmission power of the $m$-th CUE \\
			$h_{k}^{(m)}$ & link gain from transmitter in group $k$ to BS on channel $m$ &    $\mathbf{p}_k$ & vector with elements $\{ p_{k}^{(m)} \}_{m = 1, \dots, m}$ \\ \hline
				\end{tabular}
		\end{table*}
	
	\section{Channel assignment: a matching theory approach}
	\label{sec:ResourceAllocation}
	
	\begin{figure}[t]
		\centering
		\includegraphics[width=7cm,height=5cm]{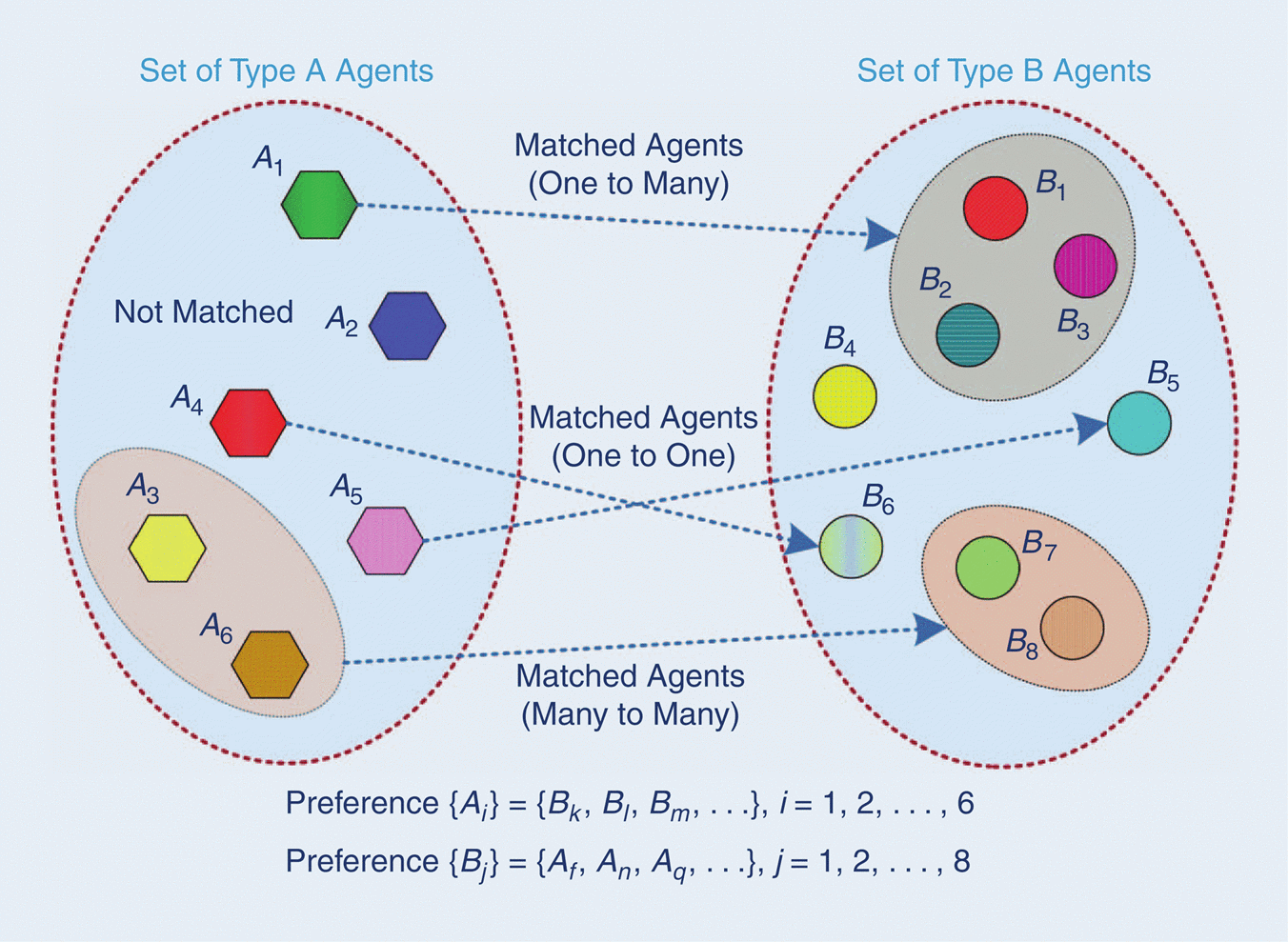}
		\caption{\label{fig:match} Matching Scenarios~\cite{Bayat2016}.}
	\end{figure}
	
	We follow a two-phase decomposition approach to solve sub-optimally the EE
	maximization problems: at the first stage, we find a (sub-optimal, in general)
	feasible allocation of channels to D2MD groups; next, in the second stage, we
	compute the optimal power allocations for each of the transmitters in the
	clusters. To solve the first sub-problem, we use the framework of matching
	theory. This Section presents our proposed solution to the channel assignment
	problem.

Matching theory considers explicitly the incentives that lead to build
mutually beneficial relations among rational and selfish
agents~\cite{Gu2015,Bayat2016}, whose preferences are given by an arbitrary
function. The outcome of the matching process, i.e., the pairing between D2MD
groups and RBs in our case, is stable, and accurately reflects the system's
objectives. While matching theory, in general, cannot
guarantee an optimal channel allocation pairing for the EE maximization, we
will choose a preference function for the matches that is based on the
aggregate interference level measured at the receivers. The rationale is
that less interference implies less transmission power for the same rate
and, since the rate is approximately linear in the SINR in the low and
medium SINR regimes, the matches will be close to optimal. Our numerical
results below support this claim.  Fig. 1  illustrates the
different scenarios that matching theory can address where D2MD groups are
elements of set $A$ and CUE users are labeled as set $B$. Clearly, the
scenario $1$ previously discussed appears in one-to-one matches, whereas
scenarios $2$, $3$ and $4$ arise in one-to-many and many-to-many matches,
respectively. We briefly review these different settings.

\subsection{One-to-One Matching}

In one-to-one matching, the objective is to find a mapping between every
element in group~$\mathcal{K}$ and an element of group $\mathcal{M}$, such
that the pairing of elements is stable, i.e., there do not exist two pairs
$(x_1, y_1)$ and $(x_2, y_2)$ such that the objective function increases by
swapping the images, namely $(x_1, y_2)$ and $(x_2, y_1)$. In our setting,
group $\mathcal{K}$ is the set of D2MD transmitters, group $\mathcal{M}$ is
the set of RBs and the goal is to identify a match that causes the minimum
possible interference. Specifically, the preference order relationship is
defined as follows.

Given two disjoint sets $\mathcal{K}$ for D2MD groups and $\mathcal{M}$ for
RBs, a one-to-one match $\mu$ is defined as a mapping from
$\mathcal{K} \cup \mathcal{M}$ to $\mathcal{K} \cup \mathcal{M}$ such that
for any $k \in \mathcal{K}$, if $\mu(k) \neq k$ then
$\mu(k) \in \mathcal{M}$, and if $\mu(m) \neq m$ for some
$m \in \mathcal{M}$ then $\mu(m) \in \mathcal{K}$. The partner $k$ referred
to as $\mu(m)$ if $\mu(m) = k $. In addition,
\begin{enumerate}
	\item $k$ prefers $m_{i}$ to $m^\prime_{i}$, if
	$\alpha_{k,m_{i}} < \alpha_{k,m\prime_{i}}$, denoted by
	$ m_{i} \prec_{\mu(k)} m^\prime_{i}$,
	
	\item $m$ prefers $k_{i}$ to $k^\prime_{i}$, if
	$\Gamma_{k_{i}} < \Gamma_{k^\prime_{i}}$, denoted by
	$ k_{i} \prec_{\mu(m)} k\prime_{i}$.
\end{enumerate}	 
The received interference on D2MD groups for $k = 1, \dots, K$ is
\begin{equation}
\label{eq:d2d-interference2}
\alpha_{k}^{(m)} =\max p^{(m)} \beta_{k,r}^{(m)} + \sum_{j \neq k}
\delta_{j,m} p_{j}^{(m)} h_{j,r}^{(m)},
\end{equation}
and  for CUE $ m = 1, \dots, M$ is
\begin{equation}
\label{eq:cue-interference2}
\Gamma_{m} = \sum_k \delta_{k,m} p_{k}^{(m)} h_{k}^{(m)}.
\end{equation}

	Clearly,(12) and (13) take into account the accumulated interference on CUE and D2MD groups, respectively. Also note that the definition presents the outcome $\mu(\cdot)$ of the matching game as a two-sided preference, generally, that is, the allocation can be based on preferences posed by both the channels and the transmitters in the D2MD groups. The definition includes the case where a channel could be empty
	($\mu(m) = m$) or a D2MD group would be forbidden to transmit on any channel
	($\mu(k) = k$), for completeness and for theoretical reasons. For application
	to a concrete case, the preference list of $k$ over $\mathcal{M}$, denoted by
	$\mathsf{D2MD}_\text{PL}$, is ranked by ascending order and similarly for CUE,
	$\mathsf{CUE}_\text{PL}$, as shown in the example of
	Table III. The central entity then executes the
	Gale-Shapely algorithm~\cite{Bayat2016} to obtain a stable match.
	Initially, group $\text{D2MD}_1$ is allocated to channel $m_{4}$, the first in
	its preference list. However, for $\text{D2MD}_2$ we notice that it also
	prefers $m_{4}$. At this step, the central entity checks $\text{CUE}_4$
	preference list, and finds out that $\text{D2MD}_2$ causes less interference
	than $\text{D2MD}_1$. Thus, swapping takes a place and $\text{D2MD}_1$ will
	never be matched to $m_{4}$ again. The algorithm continues similarly until all
	groups are matched. The final outcome is
	$\{(\text{D2MD}_1,m_3), (\text{D2MD}_2, m_1), (\text{D2MD}_3, m_4),
	(\text{D2MD}_4, m_2) \}$.

	\begin{table*}[t]
		\begin{minipage}[t]{.5\linewidth}
			\centering
			\caption{\label{table:preference} D2MD, CUE Preference Matrices.}
			\begin{tabular}{lclc}
				\textbf{Groups} & $\mathbf{D2MD}_\text{PL}$ & \textbf{CUE} & $\mathbf{CUE}_\text{PL}$ \\ \hline     
				$\text{D2MD}_{1}$ & $m_{4},m_{1},m_{3},m_{2}$ & $\text{CUE}_{1}$ & $k_{3},k_{4},k_{2},k_{1}$ \\ 
				$\text{D2MD}_{2}$ & $m_{4},m_{1},m_{2},m_{3}$ &  $\text{CUE}_{2}$ & $k_{4},k_{3},k_{1},k_{2}$ \\ 
				$\text{D2MD}_{3}$ & $m_{4},m_{2},m_{1},m_{3}$ & $\text{CUE}_{3}$ & $k_{4},k_{3},k_{1},k_{2}$ \\ 
				$\text{D2MD}_{4}$ & $m_{4},m_{2},m_{1},m_{3}$ &   $\text{CUE}_{4}$ & $k_{3},k_{4},k_{2},k_{1}$ \\ \hline
			\end{tabular}
		\end{minipage}
		\begin{minipage}[t]{.5\linewidth}
			\centering
			\caption{\label{table:preference2} D2MD, CUE Preference Matrices 2.}
			\begin{tabular}{lclc}
				\textbf{Groups} & $\mathbf{D2MD}_\text{PL}$ & \textbf{CUE} & $\mathbf{CUE}_\text{PL}$ \\ \hline
				$\text{D2MD}_{1}$ & $m_{2},m_{1}$  &  $\text{CUE}_{1}$ & $k_{3},k_{1},k_{2},k_{4}$    \\ 
				$\text{D2MD}_{2}$ & $m_{2},m_{1}$ & $\text{CUE}_{2}$ & $k_{4},k_{3},k_{1},k_{2}$ \\
				$\text{D2MD}_{3}$ & $m_{1},m_{2}$  \\ 
				$\text{D2MD}_{4}$ & $m_{1},m_{2}$  \\ \hline
			\end{tabular}
		\end{minipage} 
	\end{table*}
\subsection{Many-to-One Matching}

Many-to-one matching arises naturally in our setting when a group of D2MD
clusters are competing to use one of the available cellular resource blocks
(type 2). The reuse factor $r$ limits the maximum number of clusters per
RB. Differently to the one-to-one case, the choice of a D2MD group $k$ for a
CUE $m$ can change depending on the accumulated interference from $m$ and
other groups $k_{j}$ already allocated to $m$. The same happens to CUE $m$,
for which the aggregated interference created by a group $k$ might affect the
acceptance or rejection of a new D2MD. In detail, the preference relation is
now defined in terms of the aggregated interference as follows:	

Given two disjoint sets $\mathcal{K}$ for D2MD groups and $\mathcal{M}$ for
RBs, a many-to-one match $\mu$ is defined as a mapping from
$\mathcal{K} \cup \mathcal{M}$ to $\mathcal{K}\cup \mathcal{M}$ such that,
for $k \in \mathcal{K}$, if $\mu(k) \neq k$, then $\mu(k) \in \mathcal{M}$
and if $\mu(m) \neq m$ for some $m \in \mathcal{M}$, then
$\mu(m) \in \mathcal{K}$. The partner $k$ referred to as $\mu(m)$ if
$\mu(m) = k $.
\begin{enumerate}
	\item $k$ prefers $m_{i}$ to $m^\prime_{i}$, if $\alpha_{k,m_{i}} <
	\alpha_{k,m^\prime_{i}}$, this is denoted by $ m_{i} \prec_{\mu(k)} m^\prime_{i}$,
	
	\item $m$ prefers $k_{i}$ to $k^\prime_{i}$, if $\Gamma_{k_{i}} <
	\Gamma_{k^\prime_{i}}$, denoted by $ k_{i} \prec_{\mu(m)} k^\prime_{i}$,
	
	\item $|\mu(m)|$ $ \leq  r $, where $r$ is $m$ reuse factor,
\end{enumerate}

 where $\alpha_{k,m_{i}}$ and $\Gamma_{k_{i}}$ are 
calculated from (12) and (13). 
Note that condition 3) is simply a
restatement of constraint (10g), so it provides one connection
between the matching game and the power control subproblem.  To compute the
match, the central entity executes Gale-Shapely algorithm on several
rounds, where a single D2MD group is allocated per round, i.e., reuse factor
$r = 1$ at first (Algorithm 1). So, the central entity
first calculates interference as if it would in a one-to-one match (lines
1--4). The first round will result in some groups that will be less preferred
by all the channels in their preference lists, and the central entity will
allocate those groups in the ensuing rounds. Consider the example illustrated
in Table IV with reuse factor $r = 2$. Notice that the
number of CUEs is set to $2$, half the amount in the one-to-one case, while
the number of groups is still the same $4$. In round one (lines 9 - 10), group
$\text{D2MD}_1$ will be allocated to $m_2$. However, group $\text{D2MD}_{2}$
has identical preference as $\text{D2MD}_1$ thus, the central entity will
check $\text{CUE}_{2}$ preference list (line 11). Obviously, $\text{D2MD}_1$
causes less interference to $\text{CUE}_{2}$ so no swap action and than
$\text{D2MD}_{2}$ will be assigned to its next preference,
i.e., $\text{CUE}_{1}$. However, following the same logic for allocating
$\text{D2MD}_{3}$ to $\text{CUE}_{1}$, a swap action will take a place and
$\text{CUE}_{1}$ will be assigned to $\text{D2MD}_{3}$. By the end of round
one the outcome is a match $\{ (\text{D2MD}_3, m_1), \text{D2MD}_4, m_2)
\}$. Based on the obtained results, the algorithm re-calculates interference
(line 25) to identify how much its level increases on channel $m$ for the
allocated group when any of the free groups, i.e., $\text{D2MD}_1$ and
$\text{D2MD}_4$, are assigned. In a similar way, the algorithm calculates the
amount of interference in a free group due to such combination. The goal is to
execute the matching algorithm based on two-side preferences, so that users can be
rearranged in groups causing the minimum mutual interference to each other.

\subsection{Many-to-Many Matching}

Many-to-many matching is required to deal with the general case of splitting
and sharing. This can be seen from a resource management point of view as a
user $k$ aiming to use a set of $m$ channels, and a RB $m$ that can be
assigned to $k$ groups. The partnership relation $\mu$ can be defined in this
case as follows knowing that $\Gamma_{k_{i}}$ and $\alpha_{k,m_{i}} $ are
calculated from (13) and (12).

 Given two disjoint sets $\mathcal{K}$ for D2MD groups and $\mathcal{M}$ for
RBs, a many-to-many match $\mu$ is defined as a deterministic mapping
from $\mathcal{K} \cup \mathcal{M}$ to $\mathcal{K} \cup \mathcal{M}$ such
that, for $k \in \mathcal{K}$ if $\mu(k) \neq k$, then
$\mu(k) \in \mathcal{M}$ and if $\mu(m) \neq m$ for $m \in \mathcal{M}$,
then $\mu(m) \in \mathcal{K}$. The partner $k$ referred to as $\mu(m)$ if
$\mu(m) = k $.
\begin{enumerate}
	\item $k$ prefers $m_{i}$ to $m^\prime_{i}$, if
	$\alpha_{k,m_{i}} < \alpha_{k,m^\prime_{i}}$, denoted by
	$m_{i} \prec_{\mu(k)} m^\prime_{i}$,
	\item $m$ prefers $k_{i}$ to $k^\prime_{i}$, if
	$\Gamma_{k_{i}} < \Gamma_{k^\prime_{i}}$, denoted by
	$k_{i} \prec_{\mu(m)} k^\prime_{i}$,
	
	\item $|\mu(m)|$ $  \leq r $, where $r$ is $m$ reuse factor,
	
	\item $|\mu(k)|$ $ \leq s $, where $s$ is $k$ split factor.
\end{enumerate}

Again, conditions 3) and 4) above are the
constraints (10g) and (10f), respectively, and determine
how the preference relationship is bound to the physical resources.  The
match is found in a similar way to the one-to-many case, but with a small
difference (Algorithm 1). A small condition is added (in
italic) to fulfill the split factor. The central entity will allocate the D2MD
group to its favourite $s$ RBs. The main reason for choosing to start by the
CUEs and then moving to groups, is to ensure that a set of CUEs does not
interfere with each other. In the internal loop (lines 8--23), we guarantee the
constraint of the split factor, where a group $k$ will receive a set of not
more than $s$ RBs. Looking at the example of Table III with
$s = r = 2$, $\text{D2MD}_1$ will be allocated to $m_4$ and $m_1$, forcing the
group out of the unmatched groups list. However, the allocation remains open
to any possible update, implying that if a different $\text{D2MD}$ group is
willing to use some of its RBs and the second group is more favorable, a swap
takes place. This case appears (line 10) when we try to allocate
$\text{D2MD}_{2}$, which has similar preferences as $\text{D2MD}_{1}$, but
clearly for $\text{CUE}_{4}$ group $\text{D2MD}_{2}$ is more preferred than
group $\text{D2MD}_{1}$. Thus, the group chosen first is removed and sent back
to the unmatched group list. By the end of round one, we will have
$\{(\text{D2MD}_1, m_3), (\text{D2MD}_3, m_4), (\text{D2MD}_4, m_2, m_1) \}$,
while the remaining groups are considered for the next round to share a RB
with CUE and a D2MD. In the second round, the BS updates
$\{ \text{D2MD}_1, \text{D2MD}_2, \text{D2MD}_3 \}$ according to their
preference lists (line 22). Then, the reuse factor is increased (line 24). As
the mathematical model states, we allow groups to distribute their total rate
and maximum transmission power over the allocated resources rather than
enforcing them to equally divide those values. We remark that when a group is
replaced by another due to CUE preference, it cannot be reallocated to the
same CUE in the same round. The algorithm is a mix between one-to-many and
one-to-one where we focus on the accumulated mutual interference to decide the
candidate match.

\subsection{Stability}
The matching algorithms ensures that a CUE is allocated to its most preferred
groups. This means that the preference of a D2MD group does not force us to
assign a RB while there is a possibility for better pairing. This two-side
preference guarantees that the matching algorithm converges to a stable
solution. The partnership relation $\mu$ is stable if the pair
($k_{i},m_{i} $) does not form a blocking pair which means
$ k_{i} \prec_{\mu(k)} m_{i}$, $ m_{i} \prec_{\mu(m)} k_{i}$ for any
$k_{i} \in K$ and any $m_{i} \in M$ that are not matched with each other.  To
prove the stability we need to show that these two conditions cannot hold
simultaneously. Assume that $ m_{i} \prec_{\mu(k)} k_{i}$, thus a request had
been sent to $m_{i}$ based on preference relation $\mu(k)$ as the matching
process implies. According to this, $\mu(k_{i}) \neq m_{i}$ as $k_{i}$ is less
preferred by $m_{i}$ based on the relation $\mu(m)$. This shows that even
though $m_{i}$ is $k_{i}$'s favourite partner, yet $m_{i}$ is not interesting
in being matched with $k_{i}$. Thus, the condition
$ k_{i} \prec_{\mu(k_{i})} m_{i}$ does not hold. Similarly, the second
condition $ m_{i} \prec_{\mu(m_{i})} k_{i}$ can be proved and thus, the pair
($k_{i},m_{i} $) cannot be a blocking pair for $\mu$ which proves that such a
relationship is stable. However, in terms of optimality, there is an argument
that this match can be one-side optimal as indicated
in~\cite{Bayat2016}. Thus, we compare our results with a greedy algorithm. This
is not a problem for the power control algorithm as it is able to find the 
optimal values. For more details we refer the reader to~\cite{Hmila2017}.

\begin{algorithm}
	
	\begin{algorithmic}[1]
		\State Set up preference lists $D2MD_{PL}$, $\forall k \in K $ from
		(\ref{eq:d2d-interference2}), and $CUE_{PL}$, $\forall m \in M $,  from
		(\ref{eq:cue-interference2}) 
		\State Initial unmatched groups list \{$k_1,\dots,k_K$\}
		\State Free groups list $= \emptyset$ 
		\State Set reuse factor = 1 
		\While	{ channel capacity $< r$} 
		
		\While {unmatched groups list $\neq \emptyset$ }  
		\Repeat 
		\If { $m$ capacity $< r$}
		\State Allocate $m$ to $k$
		\ElsIf {$m$ is allocated to $k^\prime$, and $k$ is more preferred than $k^\prime$}
		\State Reject $k^\prime$ and keep $k$; send $k^\prime$ to the unmatched
		group list
		\Else 
		\State Keep $k^\prime$ and reject $k$
		\EndIf
		\If { $k$ is assigned to  $s$ RB }
		\State Remove $k$ from unmatched list
		\ElsIf {$k$ is rejected by $\forall m \in M $}
		\State Include $K$ in free groups list  
		\EndIf
		\Until all $k$ preference are tested \textit{|| $s$ is satisfied}
		\EndWhile
		\State Re-calculate accumulated interference on CUE from D2D and vice versa based on previous match.
		\State unmatched groups list = free groups list 
		\State Increase reuse factor by 1  
		\EndWhile
	\end{algorithmic}
	\caption{\label{alg:ManyToMany} Matching Algorithm for D2MD Communications.}
\end{algorithm}

\section{Optimal Power Control}
\label{sec:PowerControl}

Assume that the channel assignment is fixed. Then, the objective
function (10) is a quotient between a non-concave function and a
convex function, and the feasible set $\mathcal{P}$ is convex. Thus, the
problem of finding the optimal transmission powers falls into the general
class of fractional programming problems, for which there exist efficient
mathematical tools. For a convex set $\mathcal{C} \subseteq \mathbb{R}^n$, and
two non-negative functions $f$ and $g$, a fractional program is the
optimization problem

\begin{equation}
\label{eq:fractional-prog}
\max_{\mathbf{x} \in \mathcal{C}} \frac{f(\mathbf{x})}{g(\mathbf{x})}.
\end{equation}
The key result to solve a fractional programming problem is the following
theorem\cite{Dinkelbach67},

 A point $\mathbf{x} \in \mathcal{C}$ (14) if
and only if	$\mathbf{x}^\ast = \arg \max_{\mathbf{x} \in \mathcal{C}}
\{f(\mathbf{x}^\ast) - \lambda^\ast g(\mathbf{x}^\ast) \}$,
with $\lambda^\ast$ being the unique zero of $F(\lambda) = \max_{\mathbf{x} \in \mathcal{C}} \{ f(\mathbf{x}) - \lambda g(\mathbf{x}) \}$.

This theorem states that for solving a fractional programming problem it
suffices to find the unique zero of $F(\lambda)$. An efficient, well-known
algorithm to do that is the Dinkelbach's algorithm~\cite{Dinkelbach67}
(Alg.2). Its non-trivial step is
the computation of the maximizer $\mathbf{x^\ast}$. However, if
$f(\mathbf{x})$ and $g(\mathbf{x})$ are concave and convex, respectively, the
latter calculation is a standard convex optimization sub-problem. Here, the
feasible set is convex, but the numerator of the objective function is
non-concave. For this reason, we shall use a sequential concave programming
approach, approximating the numerator by a concave function as
in~\cite{meshgi2017optimal}:

\begin{equation}
\label{eq:log-approximation}
\log_2(1 + \gamma) \geq a \log_2 \gamma + b \quad \mathrm{where} \quad a = \frac{\overline{\gamma}}{1 + \overline{\gamma}},\quad  b = \log_2(1 +
\overline{\gamma}) - \frac{\overline{\gamma}}{1 + \overline{\gamma}} 
\log_2 \overline{\gamma}.
\end{equation}
For maximizing the minimum energy-efficiency, just recall that the minimum of
set of concave functions is also a concave function. Consequently, the
generalized Dinkelbach's algorithm (Alg.3- right) can be used to find the maximizer
point. Algorithm 2 joins all the pieces
together.

\begin{algorithm}
	
	\begin{algorithmic}[1]
		
		\State $i = 0$
		\State Pick any $\mathbf{{p}_{(k,m)}^{0}},\mathbf{{p}_{(m)}^{0}} \in \mathcal{P}$.
		\Repeat
		\State $i = i + 1$
		\State Solve~\eqref{eq:gee} or~\eqref{eq:minimum-ee} with parameters $a_{k,m}^{(i)}$,
		$b_{k,m}^{(i)}$ and $a_{m}^{(i)}$,	$b_{m}^{(i)}$ 	
		\State Set $p_{k,m}^{(i)} = 2^{q_{k,m}^{(i)}}$ and $p_{m}^{(i)} = 2^{q_{m}^{(i)}}$, where $q_{k,m}^{(i)},q_{m}^{(i)} = \arg\max \tilde{\eta}_i$	
		\State Set $\tilde{\gamma}_{k,m}^{(i)} = \gamma_k(\mathbf{p}^{(i)})$
		\State Update $a_{k,m}^{(i)}$, $b_{k,m}^{(i)}$ and $a_{m}^{(i)}$,	$b_{m}^{(i)}$  with ~\eqref{eq:log-approximation}
		\Until{convergence}
	\end{algorithmic}
	\caption{\label{alg:single-gee} Network-centric Global EE Maximization for     $M \ge 1$.}
\end{algorithm}

\begin{algorithm*}
	\centering
	\parbox{.5\linewidth}{\begin{algorithmic}[1]\footnotesize
			\State $\epsilon > 0, \lambda = 0$
			\Repeat
			\State $\mathbf{x}^\ast = \arg\max_{\mathbf{x} \in C} \{ f(\mathbf{x}) - \lambda g(\mathbf{x}) \}$
			\State $F = f(\mathbf{x}^\ast) - \lambda g(\mathbf{x}^\ast)$
			\State $\lambda = f(\mathbf{x}^\ast) / g(\mathbf{x}^\ast)$
			\Until{$F \leq \epsilon$}
		\end{algorithmic}
		\label{alg:dinkelbach}}\hfil
	\parbox{.5\linewidth}{\begin{algorithmic}[1]\footnotesize
			\State $\epsilon > 0, \lambda = 0$
			\Repeat
			\State $\mathbf{x}^\ast = \arg\max_{\mathbf{x} \in C} \min_{1 \leq i \leq
				I}\{ f_i(\mathbf{x}) - \lambda g_i(\mathbf{x}) \}$
			\State $F = \min_i f_i(\mathbf{x}^\ast) - \lambda g_i(\mathbf{x}^\ast)$
			\State $\lambda = \min_i f_i(\mathbf{x}^\ast) / g_i(\mathbf{x}^\ast)$
			\Until{$F \leq \epsilon$}
		\end{algorithmic}
		\label{alg:dinkelbach-general}}
	\caption{Dinkelbach's Algorithms. (left)~Simple. (right)~Generalized.}
	\label{alg:dinkelbachs-both}
\end{algorithm*}

\section{Numerical Results}
\label{sec:simulations}

We now present the numerical results obtained with the two-stage decomposition
approach explained in previous Sections. We assume that the cellular
users are spread over the cell area according to a homogeneous Poisson
point process $\Phi$ with density $\lambda$ devices/m$^2$. The cell is circular with radius
equal to $500$ m, and a single (tagged) central entity is located at the
center. Channel coefficients for each transmitter-receiver pair are subject to independent Rayleigh fading. For a general introduction to stochastic geometry models in
wireless networks, see~\cite{Haenggi09,Baccelli09,Elsawy13}. These
models have been extensively used for modeling D2D
communications~\cite{George15, Afshang16},
for general interference modeling~\cite{Ak16a,Ak16b}, for non-orthogonal
communications~\cite{Tabassum16} and, more recently, for millimeter wave wireless
systems~\cite{Venugopal16}.

We have used two clustering algorithms for the
grouping of the D2MD receivers: 
\begin{enumerate}
	\item \textbf{$\mathbf{K}$ Nearest Neighbour (KNN)}. The $K$ heads of
	cluster (transmitters) are randomly selected among the points in $\Phi$. 
	The rest of users/points in
	$\mathcal{S}$ are considered as potential receivers assigned to the
	closest group head. Finally, only the groups that reach the target
	size $|\mathcal{S}_{k}|$ are kept.
	
	\item \textbf{Distance limited (DL)} The number of clusters $K$ is
	specified in advance, as in KNN. However, the distance between
	transmitter/receivers is explicitly controlled with a parameter
	$d_{\text{max}}$, defined as a fraction of the cell
	radius. Therefore, all users located within distance
	$d_{\text{max}}$ of some of the $K$ head clusters are retained as
	receivers. DL allows to form heterogeneous groups with unicast and
	multicast communications simultaneously.
\end{enumerate}

Table V lists the physical parameters used during our
numerical tests. In the plots reported in the rest of this Section, each point
is the average of at least $200$ independent simulation runs. Confidence
intervals for these runs have been computed, but have been omitted from the plots 
for clarity.
\begin{table*}
	\centering
	\caption{\label{table:params} System Parameters for the Numerical
		Experiments}
	\begin{tabular}{lclc} \hline
		\textbf{Parameter} & \textbf{Value}& \textbf{Parameter} & \textbf{Value}\\ \hline
		Cell radius & $500$ m  & Reuse factor ($r$)  & $[2,3,4,5]$ \\   
		Network density ($\lambda$) & $250\,$devices/m$^2$ & Minimum transmission rate & $0.1,0.5$ bi/s/Hz \\  
		Number of D2D groups & $[4,5,6,9,10,15,25]$ &     Maximum transmission powers & $[-5,25]$ dBm \\  
		Number of CUE users & $[3,4,5,6,8,10,15]$ & Noise power density ($N_0$) & $-100$ dBm/Hz \\  
		Split factor ($s$)  & $[2,3,4]$ & Circuit Power &$10$ dBm  \\  
		Path loss exponent & $2.5$  \\ \hline    
	\end{tabular}
\end{table*}

\subsection{Performance Evaluation} 

\begin{figure*}[t]
	\centering
	\includegraphics[width=9cm,height=5cm]{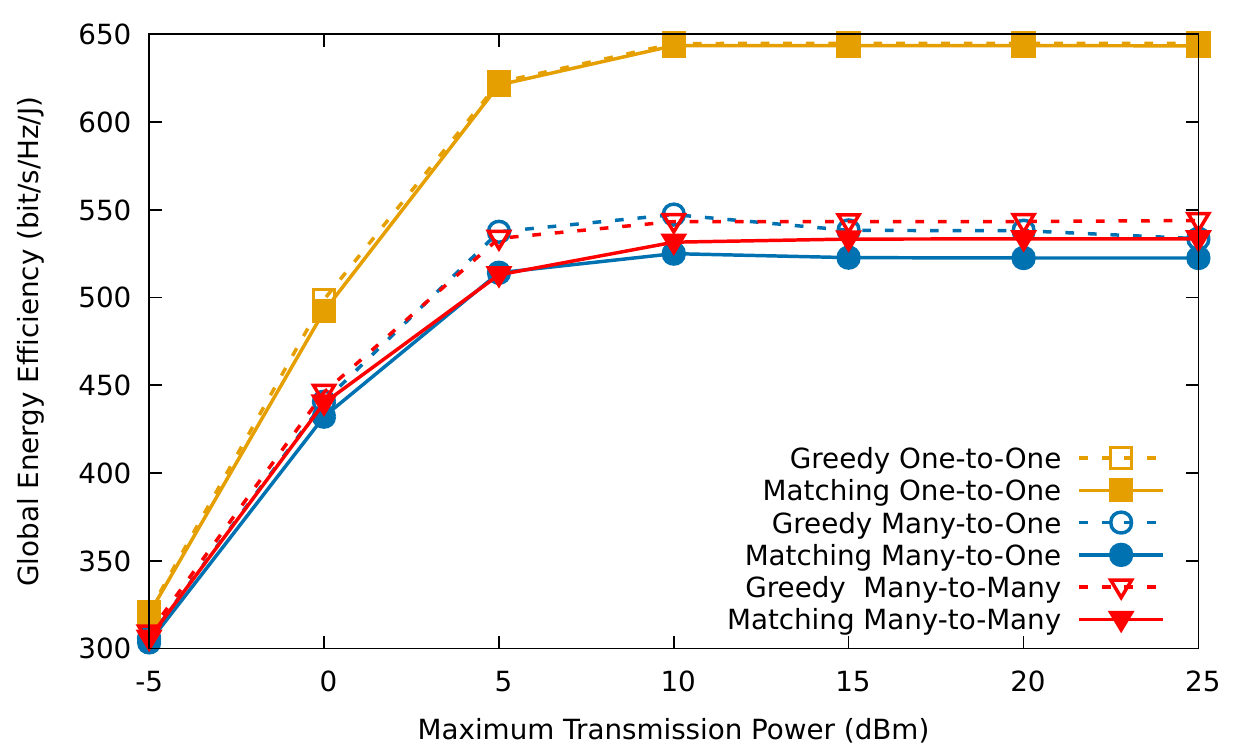}\hfil
	\includegraphics[width=9cm,height=5cm]{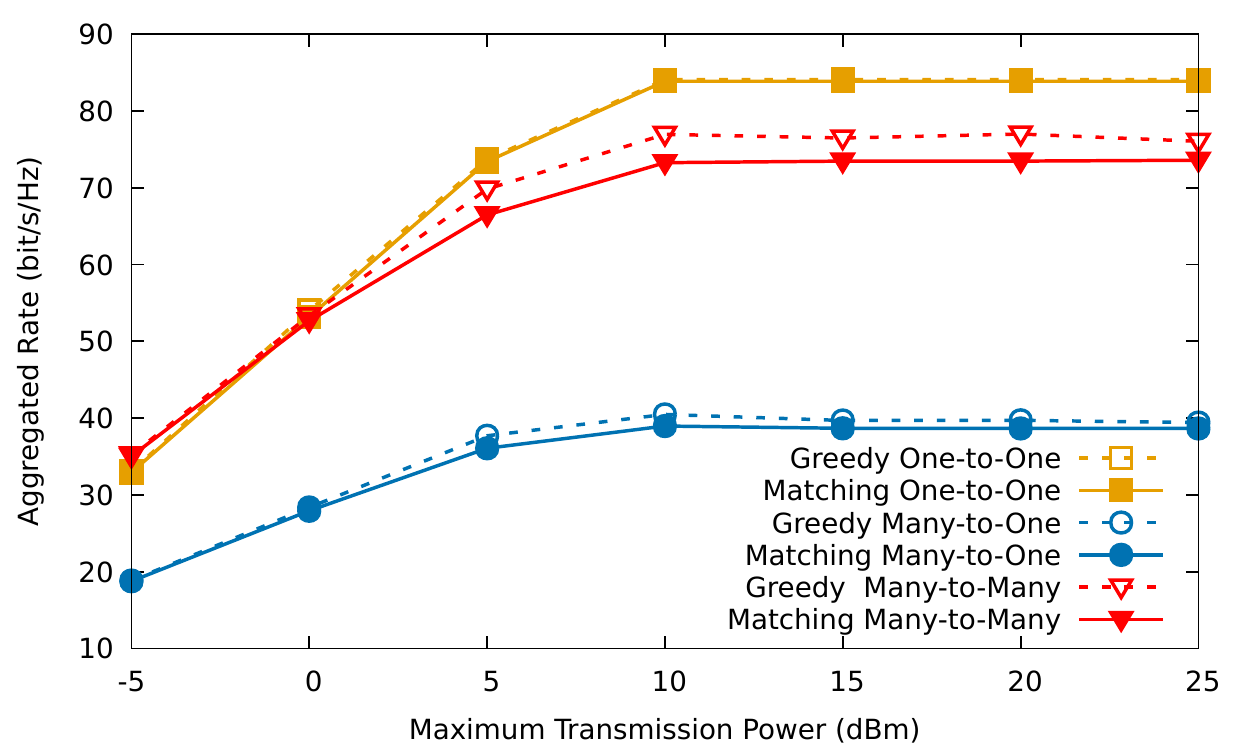}
	\caption{\label{fig:greedy} GEE and Aggregated Rate vs. Transmission Power For The Greedy and The Proposed Matching Algorithms.}
\end{figure*}

\begin{figure*}[t]
	\centering
	\includegraphics[width=9cm,height=5cm]{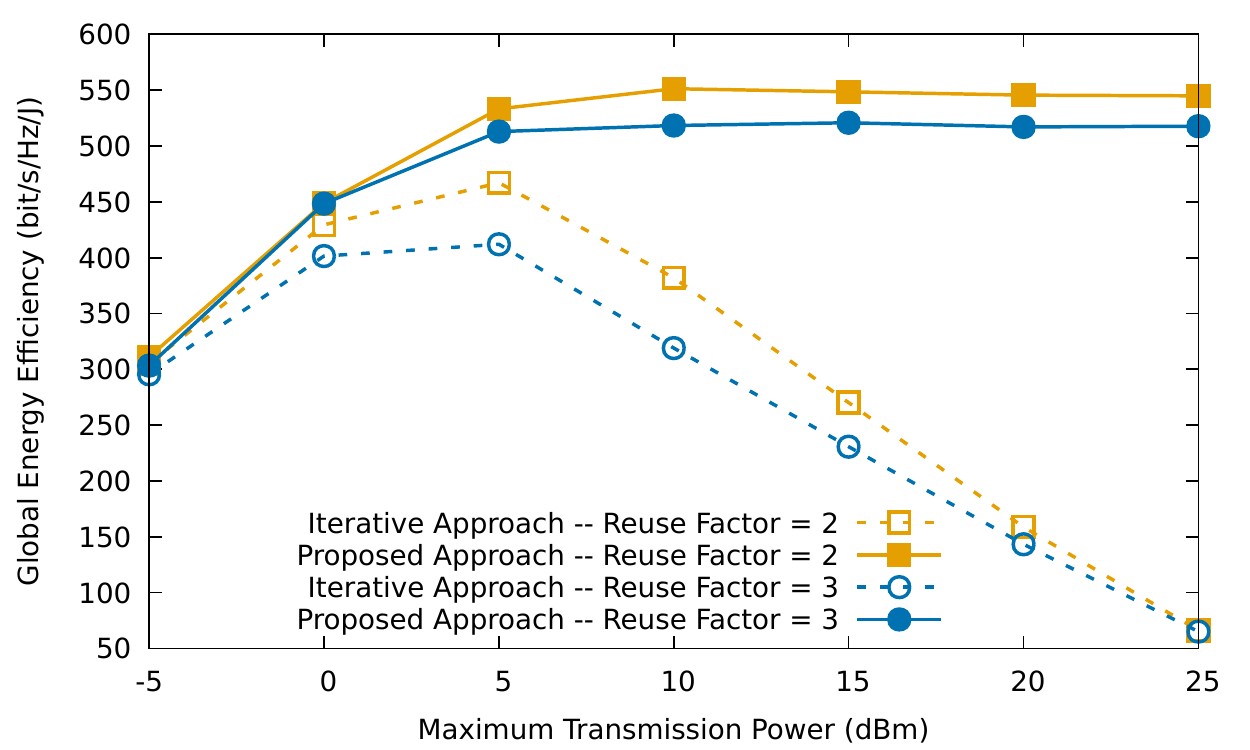}\hfil
	\includegraphics[width=9cm,height=5cm]{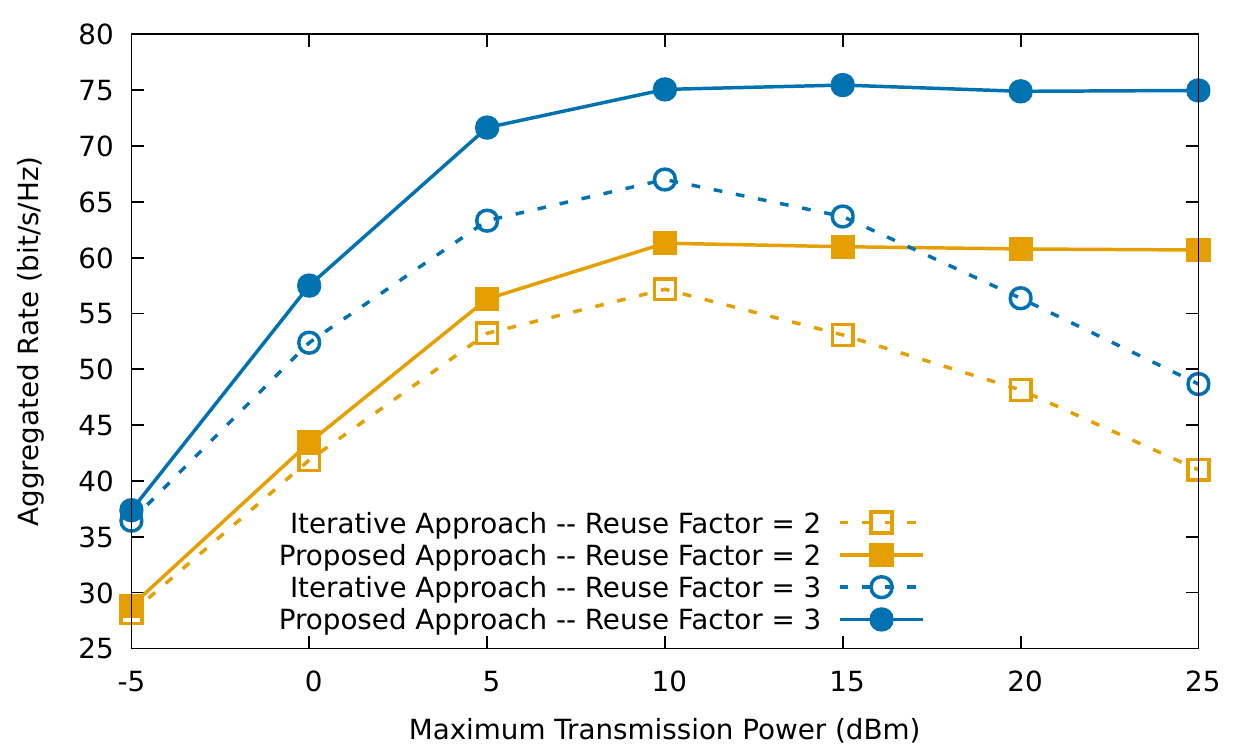}
	\caption{\label{fig:iterativeEE} GEE and Aggregated
		Rate vs. Transmission Power For The Iterative and The Proposed
		Matching Algorithms.}
\end{figure*}

\begin{figure*}[t]
	\centering
	\includegraphics[width=9cm,height=5cm]{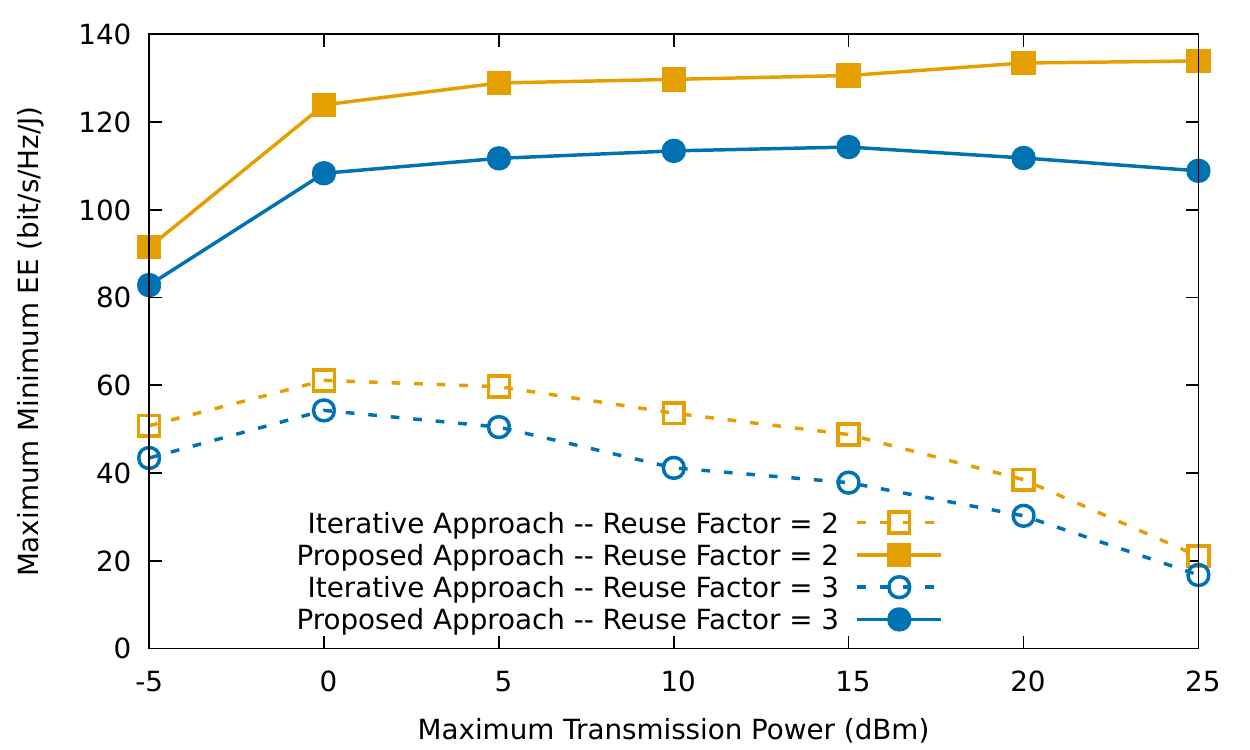}\hfil
	\includegraphics[width=9cm,height=5cm]{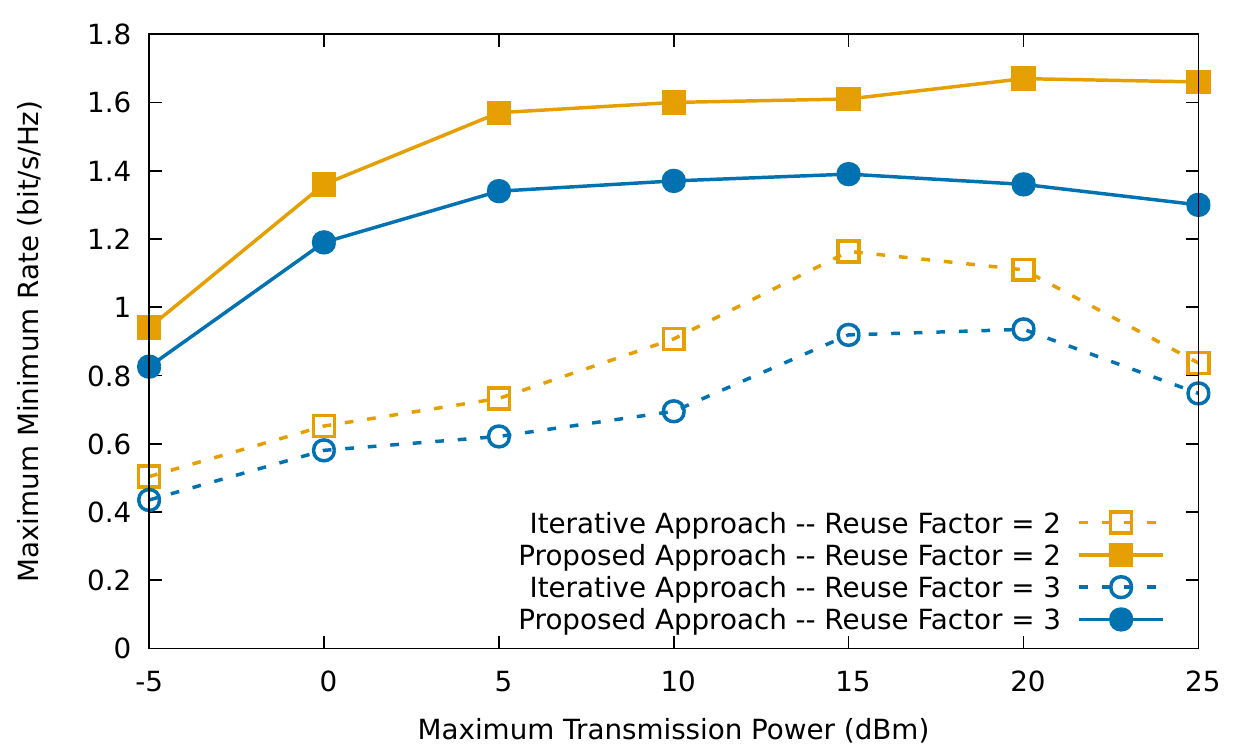}
	\caption{\label{fig:iterativfeEE} GEE and Aggregated
		Rate vs. Transmission Power For The Iterative and The Proposed
		Matching Algorithms.}
\end{figure*}

First, we compare the performance of the matching-theory-based solution to a
greedy algorithm. Differently from our approach, at each round, the greedy
algorithm explores all the possible D2MD and CUE pairings via
Algorithm 2. The CUE$^\prime$ and D2MD$^\prime$ achieving
the highest energy efficiency are selected at each step. Though a greedy
approach is not necessarily optimal, we have found that its performance is
generally very close to the optimal solution, and the running time is
substantially lower than an exhaustive search, whose complexity is
exponential.

In D2MD, the greedy algorithm can be sensitive to variable group sizes. This
is because larger clusters can achieve higher EE, which in turn may affect the
selection of optimal pairs. For this reason, we use the KNN clustering
technique in all the testing cases to guarantee a fair comparison. In the
one-to-one matching scenario, we set $5$ D2MD groups of equal size with
$3$ receivers and the minimum rate per channel is set to
0.1 bit/s/Hz. For the many-to-one matching scenario, we set
the number of D2MD users to $4$, the reuse factor $r$ is set to $2$, the
minimum rate per channel is set to 0.1 bit/s/Hz  and the
number of RBs is set to $2$. Finally, in the many-to-many case, the number of
clusters is $4$, the number of RBs is $4$, and the split and reuse factors are
set to $r = s =2$. In all the simulations, we vary the maximum transmission
power per user in the range $[-5, 25]\,$ dBm. Fig. 2
depicts, respectively, the global EE and the aggregated rate, and exhibits
that the results obtained with the proposed algorithm are close to optimal.

Secondly, we compare our results with the iterative matching
algorithm from~\cite{Zhou2016}. Similar configurations were used, but with
$M = 3$, $K = \{ 6, 9\}$, $r = \{2, 3\}$, and $s =
1$. Fig. 3  and 4  illustrate
both the system and the individual EE and rate. Clearly, EE increases
monotonically with the power budget of a user until reaching a maximum. This
happens at, approximately, the same values for our technique and
for~\cite{Zhou2016}, but the latter shows a decreasing performance beyond that
point. The graphs show that our algorithm handles better the aggregated
interference, which in turn results in a significant performance improvement
in both EE and rate. 

\subsection{One-to-One Matching} 

We use the DL clustering technique to create heterogeneous groups including
both uni- and multi-casting.  The $M$ CUE devices with the best channel
quality are automatically chosen to share their RBs with the $K$ D2MD
groups. Here, $K = M = \{ 5, 10, 15 \}$, and each group has $3$ receivers on
average. The maximum distance between the head cluster and a receiver is
approximately 50 m . The minimum rate per channel is set to
0.1 bit/s/Hz.

\subsubsection{Feasibility and Convergence Rate}
Since the location of the devices and the forming of clusters are random, it
is not guaranteed that a given scenario is feasible for a specified set of
constraints and power budget. For this reason, we always average our results
over a constant number of $200$ \emph{feasible} cases, to make comparisons
fairer.  At the same time, we kept track of the number of infeasible cases to
have a clear idea about the effect of the power budget on the problem
feasibility. Here, a feasible case is that all the users (CUE, D2MD) were
able to satisfy both the minimum rate and maximum power constraints. When the
number of D2MD users is $5$, the total number of non-feasible cases started
from $26$, and went down to $4$ as the transmission power budget was
increased. This number increases to $80$ (resp., $304$) when the number of
transmitters increases to $10$ (resp., $15$), and reduces to $12$ as more
transmission power was assigned to users during the maximization of the system
EE. Similar values were observed for maximizing the minimum EE, where the
non-feasible cases fell down to $0$, $1$, $9$ for a number of transmitters $M$
equal to $5$, $15$, $20$, respectively. As expected, we found that, either
for global EE or for max-min EE, feasibility increases with the power budget
and decreases with the number of users, potentially producing higher
interference. Nevertheless, note that because of the power optimization, the
interference level is not necessarily proportional to the number of
users. Further, we also measured (over the $200$ simulation
runs) the number of iterations needed by the matching algorithm to
converge. For $K = M = 5$, the matching algorithm converges in $3$
iterations, and for $K = M = 10 $ and $15$, it converges in $5$ and $6$
iterations, respectively. 

\subsubsection{Energy Efficiency and Rate Analysis}

The effect of varying the maximum allowed transmission power is depicted in
Fig. x, which shows the GEE in three different cases,
$K = 5, 10, 15$. Clearly, energy efficiency increases with the power allocated
to individual users, up to a peak point. This can be seen around $K = 10$,
where GEE reaches its highest value at $10$ dBm. GEE remains almost
constant beyond that point, only with small decrease. The same conclusions
hold for MEE, as shown in Fig.x . In addition, we can
clearly notice that the highest EE was achieved when the number of groups is
the lowest. This is because the amount of drawn power continuously increases
with the number of users. The average power used while maximizing the global
EE rises from $82.3$ mW up to $390.9$ mW , while for
an individual D2MD transmitter it falls between $10.3$ mW  and
$12.4$ mW . On the contrary, the global aggregated rate
continuously increases as more users coexist in the network. Again, in all the
experiments the rate reaches a saturation point where no further improvement
is possible due to the aggregated interference.

\begin{table}[t]
	\centering
	\caption{\label{table:distance_limitGEEOneToOne} EE
		(bit/s/Hz/J) and Average Rates
		(bit/s/Hz) with One-to-One Matching}
\begin{tabular}{lcccccc}\hline
	\textbf{Min. Rate} &\textbf{GEE} & \textbf{Agg. Rate} & \textbf{MEE} & \textbf{User Rate}  \\  \hline
	
	$0.1$  &  $791.42$   &   $102.39$    &  $181.62$  &  $6.04 $  \\   
	$0.2$  &  $794.41$   &   $102.29$    &  $183.14$  &  $6.33$  \\   
	$0.3$  &  $805.37$   &   $103.48$    &  $185.86$  &  $6.64$  \\   
	$0.4$  &  $791.23$   &   $101.65$    &  $179.99$  &  $6.17$  \\  
	$0.5$  &  $783.46$   &   $100.40$    &  $179.19$  &  $6.16$  \\  \hline
\end{tabular}
\end{table}

\subsubsection{Minimum Rate Constraints} 

The effect of a tighter rate constraint was also tested within the interval
$[0.1, 0.5]\,$ bit/s/Hz for the total rate. For ease of
comparison with previous cases, the same configuration is kept with
$K = M = 5$. In this test case, CUE and D2MD users are transmitting with
a maximum power equal to
$10$ dBm. Table VI shows that both GEE and
global rate decrease when the rate constraint is stronger. This is a clear
indication that setting a higher value for the total transmission rates forces
the devices to use proportionally higher power in order to satisfy the
constraint. Since rate increases only logarithmically with the SIR in the low
or moderate SIR region, these higher rates do not compensate the extra energy
expenditure.

\begin{figure*}[t]
	\centering
	\includegraphics[width=9cm,height=5cm]{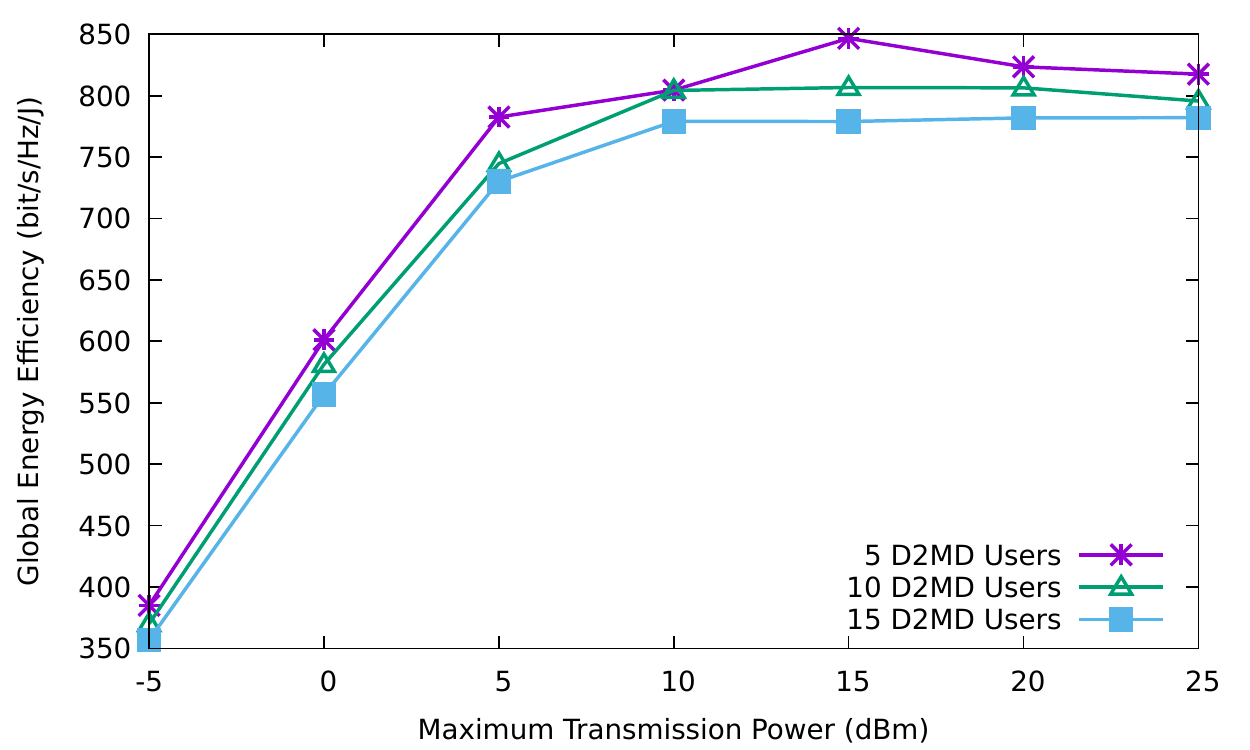}\hfil
	\includegraphics[width=9cm,height=5cm]{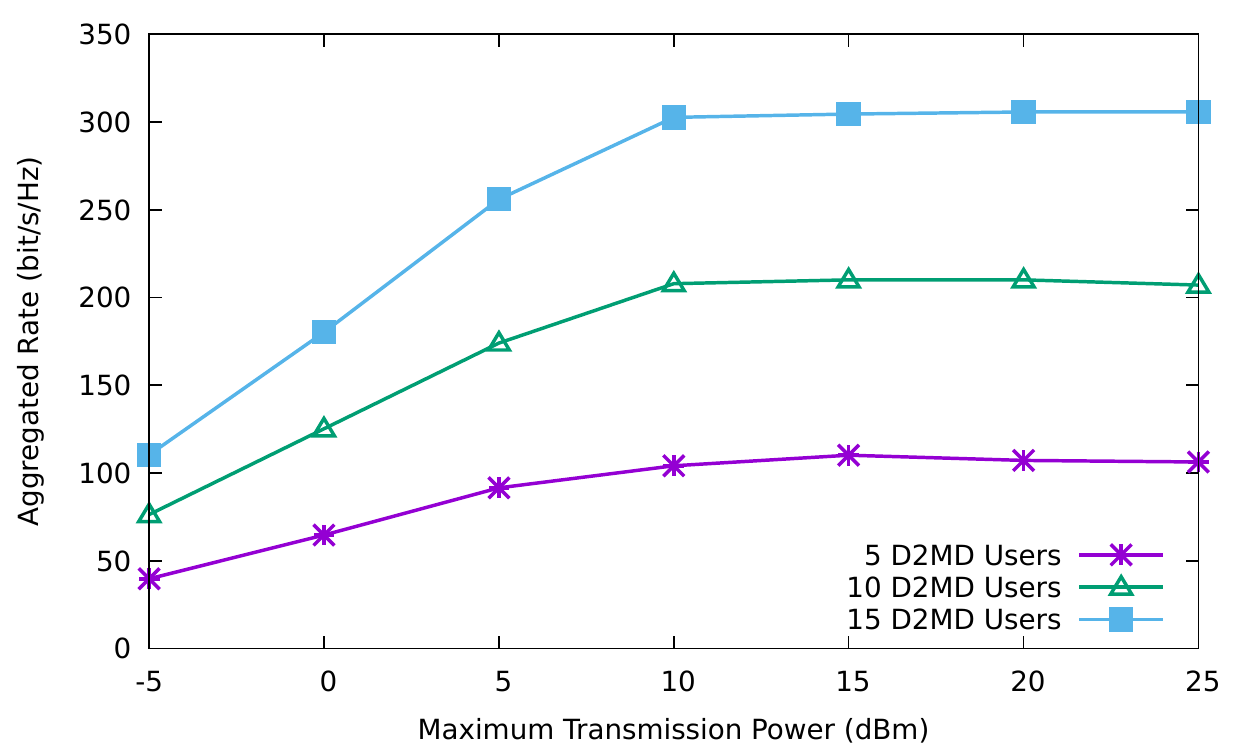}
	\caption{\label{fig:RateGEE-oneToOne} GEE and Aggregated Rate vs. Transmission Power with One-to-One Matching.}
\end{figure*}

\begin{figure*}[t]
	\centering
	\includegraphics[width=9cm,height=5cm]{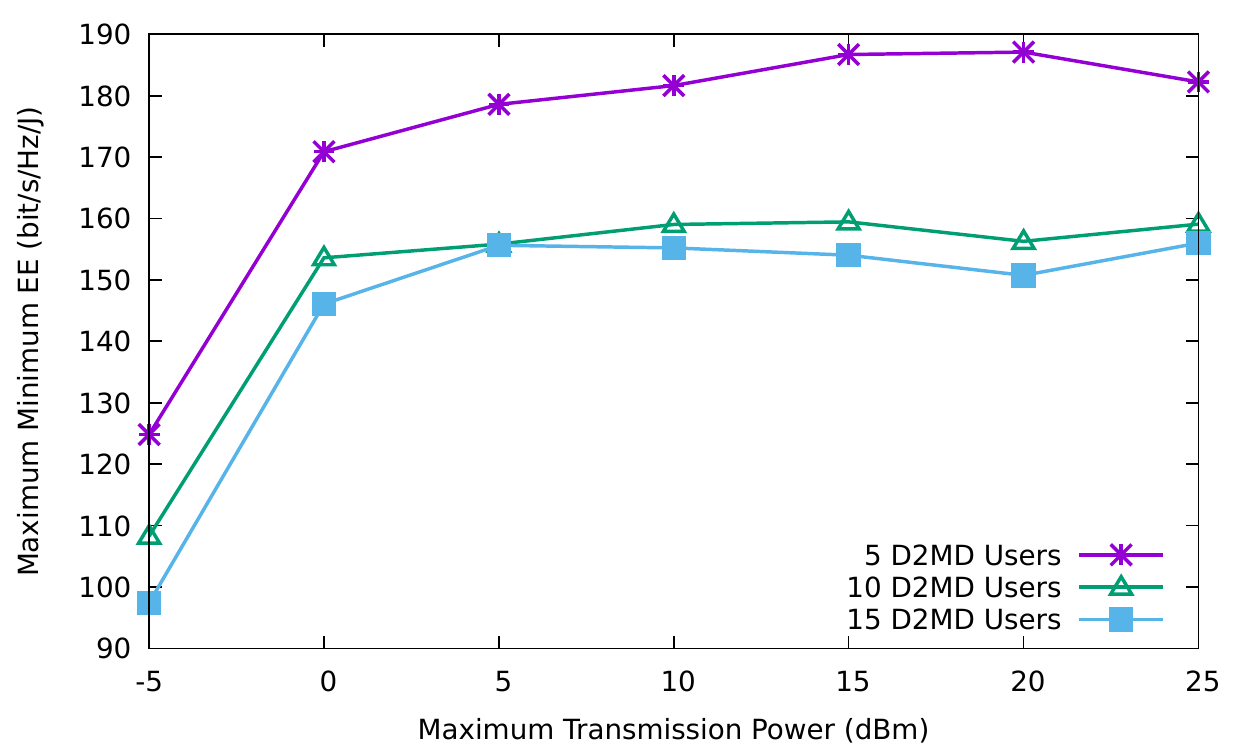}\hfil
	\includegraphics[width=9cm,height=5cm]{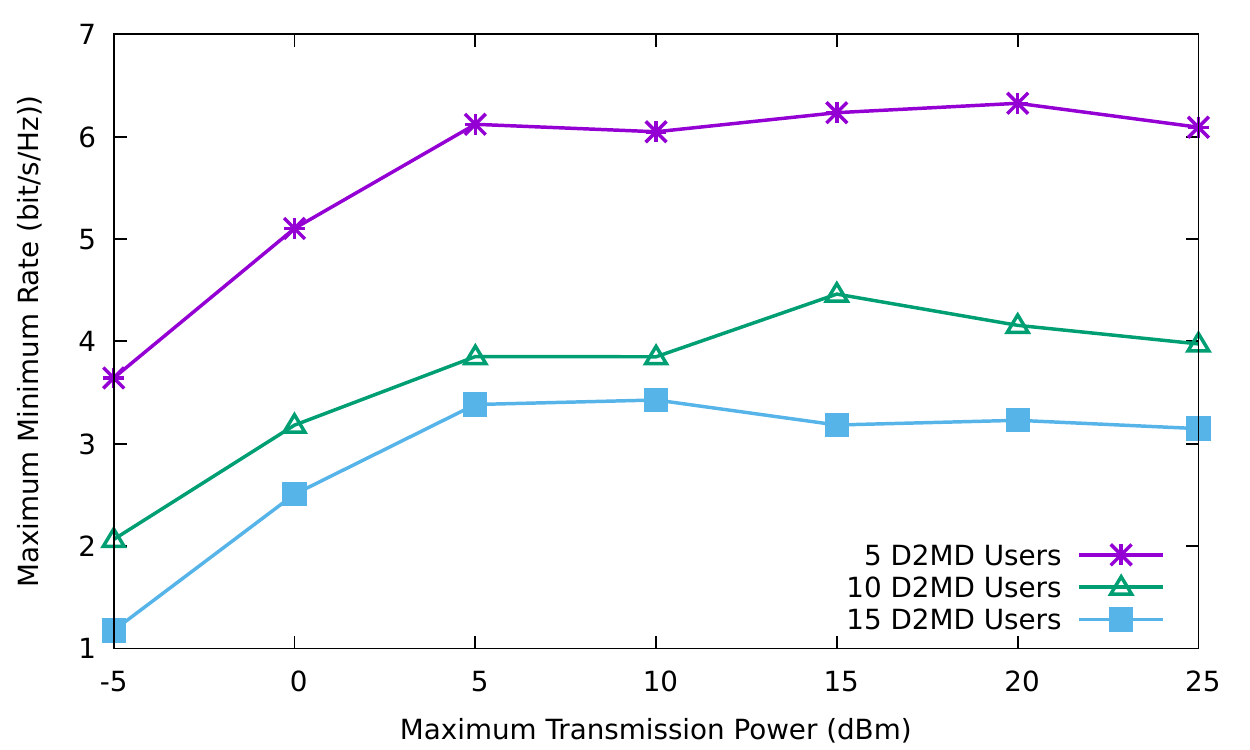}
	\caption{\label{fig:RateMEE-oneToOne} MEE and Minimum Rate vs. Transmission Power with One-to-One Matching.}
\end{figure*}

\subsection{Many-to-One Matching}

\begin{figure*}[t]
	\centering
	\includegraphics[width=9cm,height=5cm]{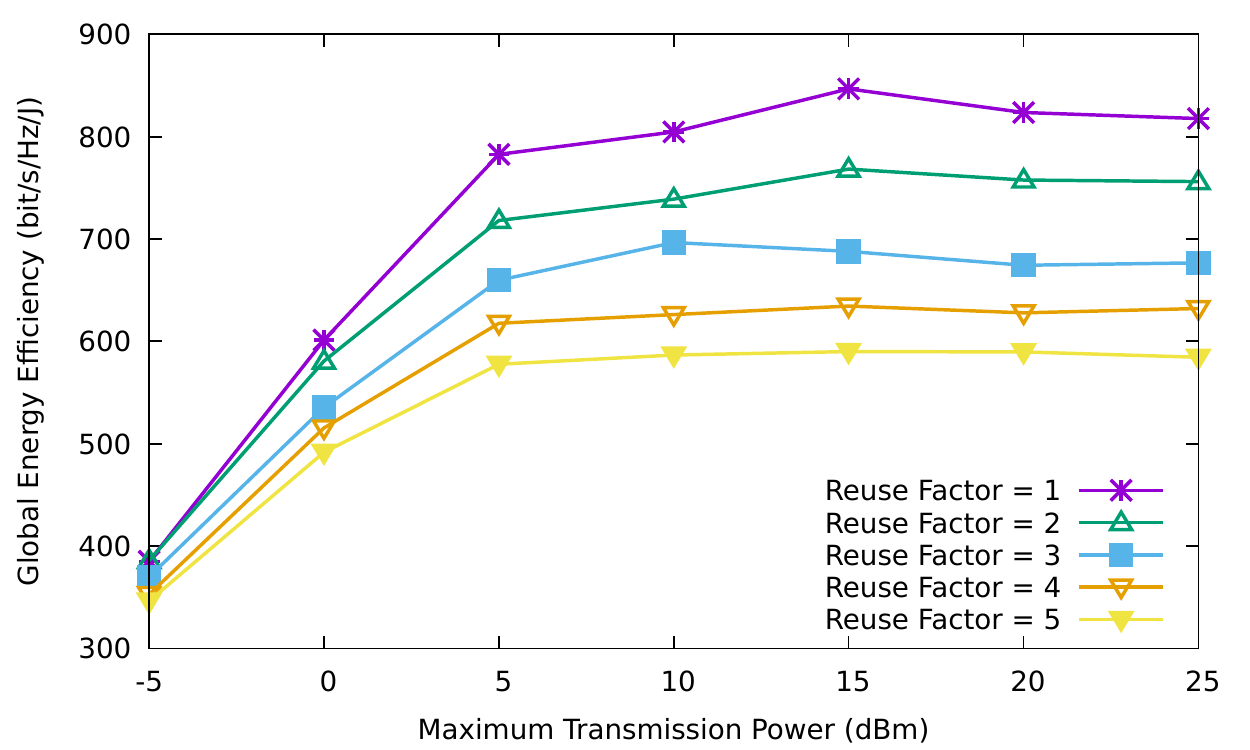}\hfil
	\includegraphics[width=9cm,height=5cm]{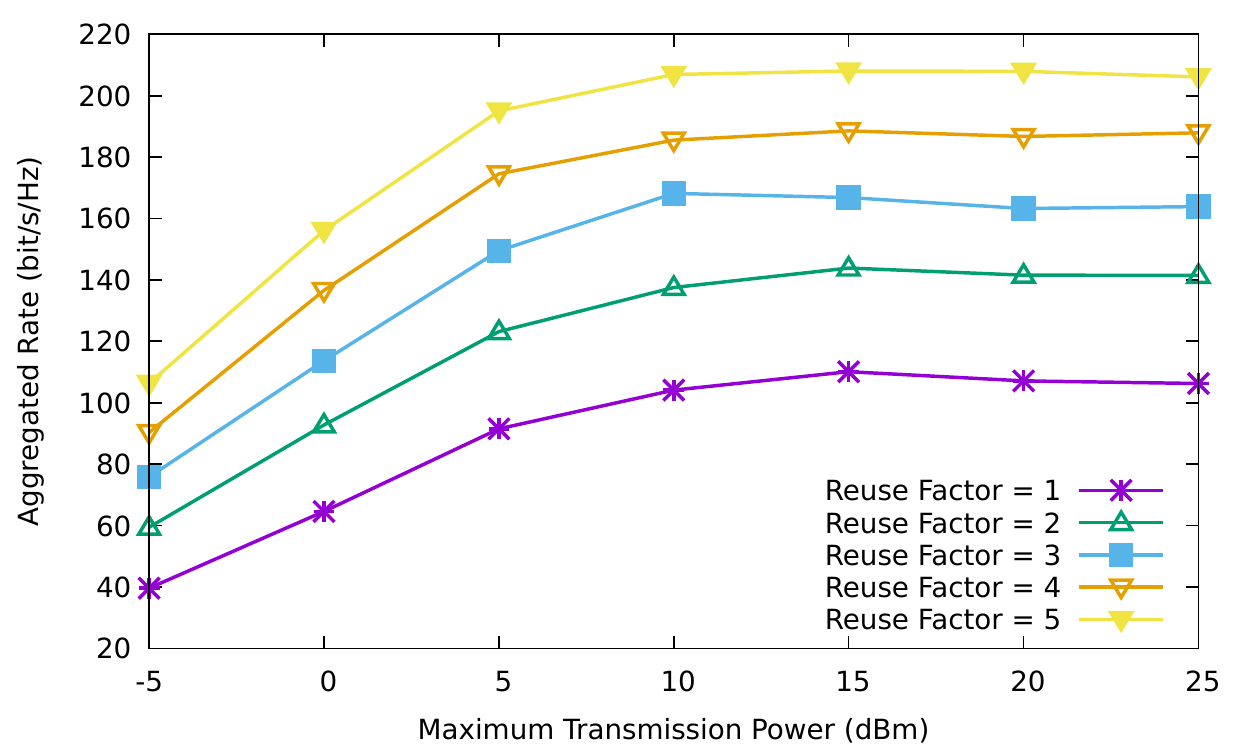}
	\caption{\label{fig:MT1-gee} GEE and Aggregated Rate vs. Transmission Power with Many-to-One Matching.}
\end{figure*}

\begin{figure*}[t]
	\centering
	\includegraphics[width=9cm,height=5cm]{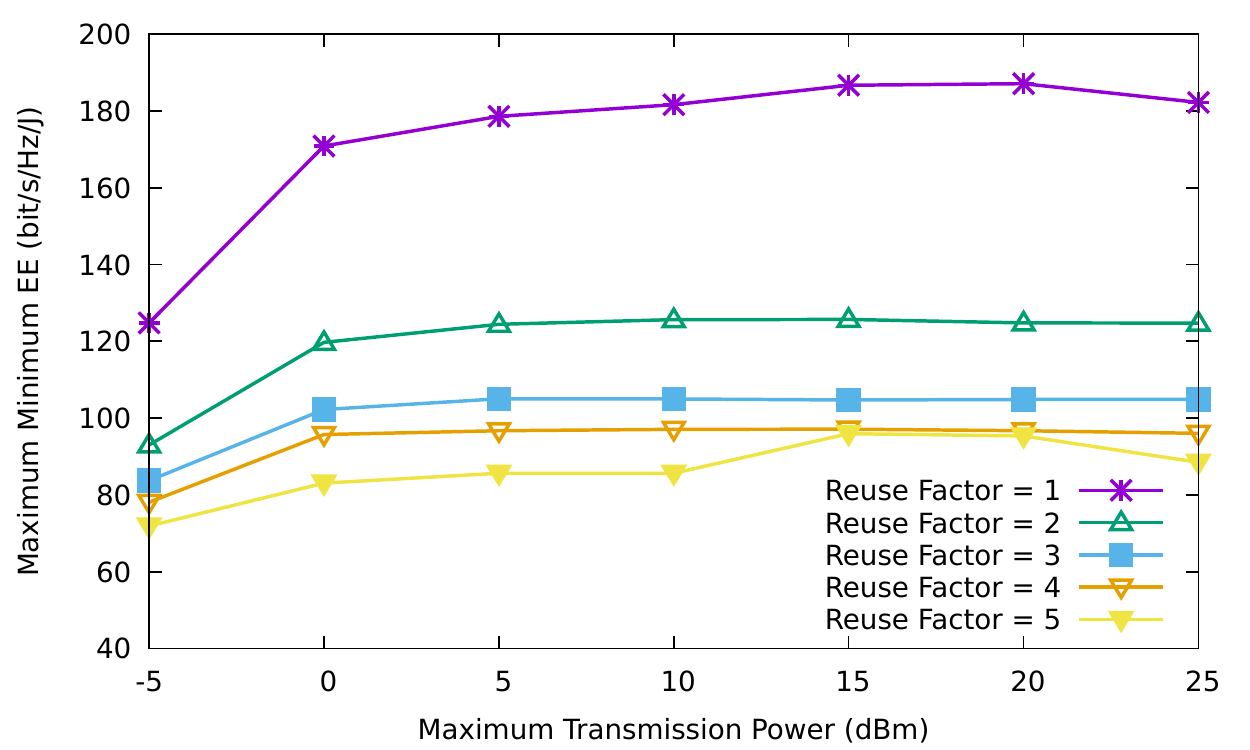}\hfill
	\includegraphics[width=9cm,height=5cm]{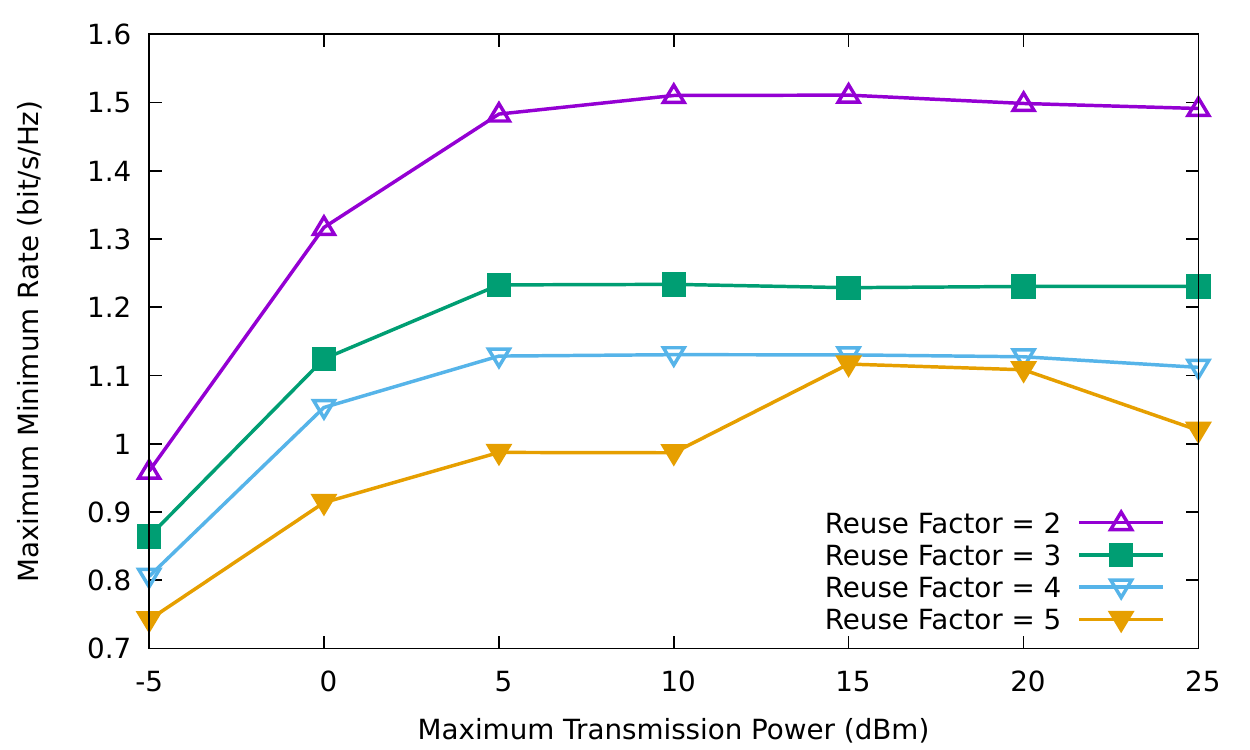}
	\caption{\label{fig:MT1-mee} MEE and Minimum Rate vs. Transmission Power with Many-to-One Matching.}
\end{figure*}

\begin{table*}[t]
	\centering
	\caption{\label{table:distance_limitGEEManyToOne} EE
		(bit/s/Hz/J) and Average Rates
		(bit/s/Hz) with Many-to-One and Many-to-Many Matchings}
	\begin{tabular}{cccccccccccc}
		& \multicolumn{4}{c}{\textbf{Many-to-One}} & 
		\multicolumn{4}{c}{\textbf{Many-to-Many}} \\
		\hline
		\textbf{Min. Rate} &\textbf{GEE} & \textbf{Agg. Rate} & \textbf{MEE} & \textbf{User Rate} & \textbf{GEE} &\textbf{Agg. Rate} & \textbf{MEE} & \textbf{User Rate}    \\ \hline

		$0.1$  &  $738.86$   &  $137.53$    & $125.67$   & $1.51$ &  $698.54$   &   $188.49$  & $120.16$ &   $1.46$  \\   
		$0.2$  &  $743.94$   &  $137.65$    & $127.81$   & $1.54$ &  $705.08$   &   $189.79$   & $121.45$   & $ 1.48$ \\ 
		$0.3$  &  $733.58$   &   $129.97$   & $128.27$   & $1.55$  &  $702.11$   &   $189.05$   & $126.55$   &  $1.56$ \\   
		$0.4$  &  $707.27$   &   $101.65$   & $128.86$   & $1.56$ &  $678.10$   &   $183.13$   &  $139.23$   &   $1.73$ \\   
		$0.5$  &  $608.10$   &   $128.03$   & $129.33$   & $1.57$ &  $678.10$   &   $183.13$   &  $122.52$   &   $ 1.49$ \\  \hline
	\end{tabular}
\end{table*}

Similarly, we applied the DL clustering algorithm to create $K$ clusters of an
average size equal to $3$. The distance ratio is $d_{max} = 1/8$, such that
the head cluster is around $50$ m away from the receivers. Using the
many-to-one matching algorithm for channel allocation, we aim to highlight the
effect of the reuse factor on both EE and rate. Thus, we fix the number of
cellular users $M = 5$ and vary the reuse factor $r$ from $1$ to $5$, for
$K \in \{ 10, 15, 20, 25\}$, respectively. Obviously, when $r=1$, the
experiment corresponds to one-to-one matching. The rest of system
parameters remain the same as listed in Table V.

\subsubsection{Feasibility and Convergence Rate}

Tracking the percentage of feasible cases is also important
with the matching-theoretic approach, since the range of allowed power
budget could not yield for a feasible solution. The problem feasibility
should become easier as the transmission power budget increases. Indeed, this
is exactly what we observed, and coincides with the observations in previous
numerical experiments. Regarding the convergence of the matching algorithm, it is again achieved in few
iterations. For example, for $M = 5$ and $K \in [10, 25]$, the algorithm
converges in $6$, $10$, $13$, and $16$ iterations, for a reuse factor $r$
equal to $2$, $3$, $4$ and $5$, respectively. 

\subsubsection{Energy Efficiency and Rate Analysis} 

As Fig. 6 illustrates, both GEE and the aggregated rate
increase with the power budget allocated to CUE and D2MD
transmitters. However, when more D2MD clusters share the same RB, the
interference becomes stronger, so users tend to use more power to guarantee
the QoS constraints. For a reuse factor $r = 2$, the total needed power rises
from $154.1$ mW up to $187.0$ mW for transmission
powers equal to $-5$ dBm and $25$ dBm, respectively. The total
employed power when $r = 5$ is $307.0$ mW and goes up to
$352.3$mW for maximum transmission powers equal to $-5$ dBm
and $25$ dBm, respectively. Thus, the total rate is higher when more
users are allocated to a RB, yet this is not the case for
GEE. Fig. 7 shows the minimum EE and the rate. Here, we see
the impact of power and $r$ on individual users. We should mention that
these values represent CUE/D2MD individual rate and EE. We notice that the
behavior of both MEE and the minimum rate is similar to GEE and the aggregated 
rate as the users' power budget increases.

\subsubsection{Minimum Rate Constraints} 
Here, we investigate the effect of a tighter rate constraint within the
interval $[0.1, 0.5]\,$ bit/s/Hz on EE and rate for the
many-to-one matching algorithm. The number of CUEs is set to $5$, while the
number of D2MD clusters is set to $10$ to satisfy the reuse factor $r = 2$.
The maximum transmission budget for CUE and D2MD users is set to
$10$ dBm. The results are illustrated in
Table VII. Similar to the one-to-one
scenario, both EE and rate decrease when the rate constraint is stronger. This
is a clear illustration of the trade-off between energy efficiency and
maximization of the sum-throughput when the system capacity has been already
reached.

\subsection{Many-to-Many Matching}

We complete our experiments by studying the influence of allowing D2MD groups
to use more resources. Here, the DL clustering algorithm is applied to create
$K$ groups of average size equal to $3$. The number of clusters is fixed to
$4$ while the available channels $M$ varies from $4, 6, 8$ with split factor
$s$ equal to $2, 3, 4$, respectively. The main reason for choosing these
parameters is to ensure that the reuse factor $r = 2$ is always
satisfied. Thus, we force the presence of inter-groups interference and
simulate the general case of resource allocation rather than a particular
version of the many-to-one matching algorithm.

\subsubsection{Feasibility and Convergence Rate} This were tested in the scenarios previously
detailed. Obviously, solutions exist for the lowest transmission power
($-5$dBm), yet the number of non-feasible cases is considerably high
(e.g., $48$, $45$ and $50$ for DL). Increasing the D2MD power budget up to
$25$ dBm significantly reduces the number of non-feasible cases down to
$[8,6,14]$ for DL. Finally, the fractional programming
algorithm 2  converges in an average of 7--8 rounds, though
this figure increases with the amount of users. Moreover, we
also considered the number of iterations needed by the matching algorithm to
match all users and to satisfy $r$ and $s$ averaged over $200$ cases. For
$K = M = 4$ and $s = r = 2$, the matching algorithm converges in $4$
iterations. Holding the same number of clusters and $r$ while setting CUE to
$6$ and $8$ and $s = 3$ and $4$, the matching algorithm converges in $4$ and
$5$ iterations, respectively. 

\begin{figure*}[t]
	\centering
	\includegraphics[width=9cm,height=5cm]{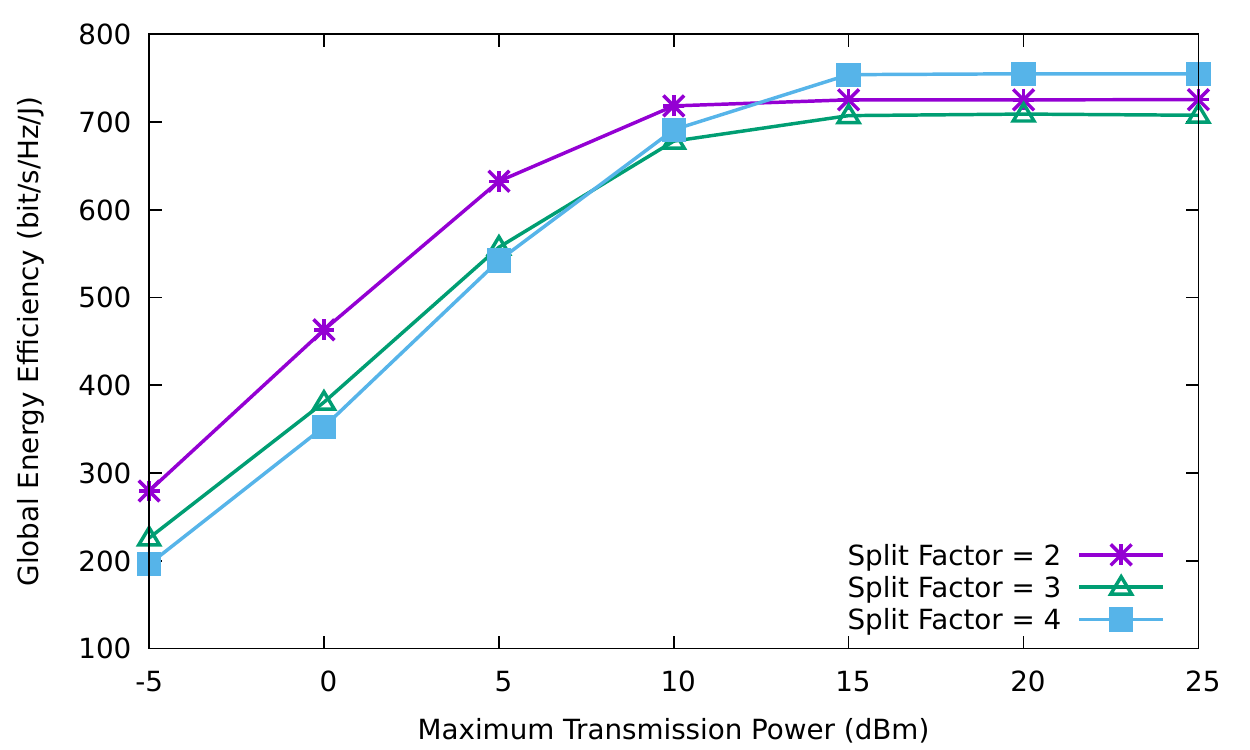}\hfil
	\includegraphics[width=9cm,height=5cm]{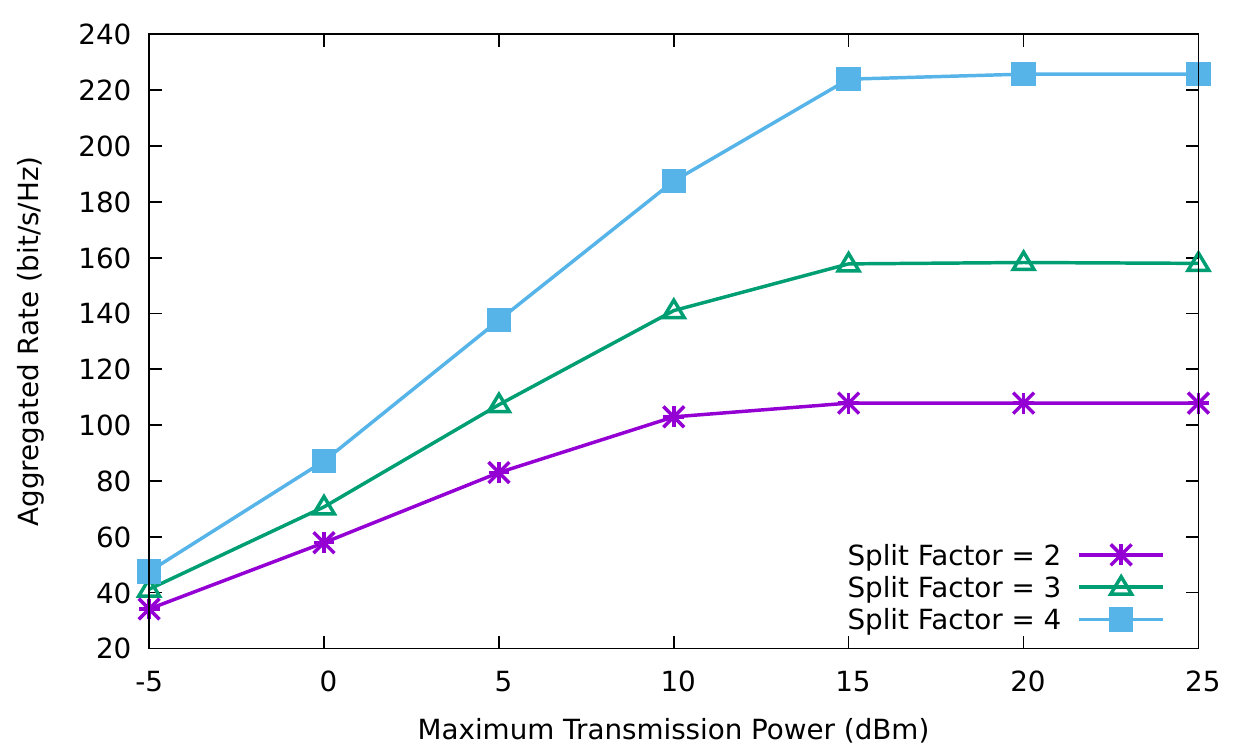}
	\caption{\label{fig:MTM-gee} GEE and Aggregated Rate vs. Transmission Power with Many-to-Many Matching.}
\end{figure*}

\subsubsection{Minimum Rate Constraints} As in the previous cases, we study
the effect of a tighter rate constraint when the total rate to be achieved
over $s$ RBs is in the range of $[0.1, 0.5]\,$ bit/s/Hz. For
this, we set the number of D2MD to $4$, CUE to $8$ and the split factor $s$ to
$4$. Moreover, to avoid the presence of extremely low throughput in any of the
allocated RBs, we set the minimum rate per channel to
$0.01$. Table VII shows that both EE and rate
decrease when the rate constraint is stronger, so the
previous conclusions hold here too.

\begin{figure*}[t]
	\centering
	\includegraphics[width=9cm,height=5cm]{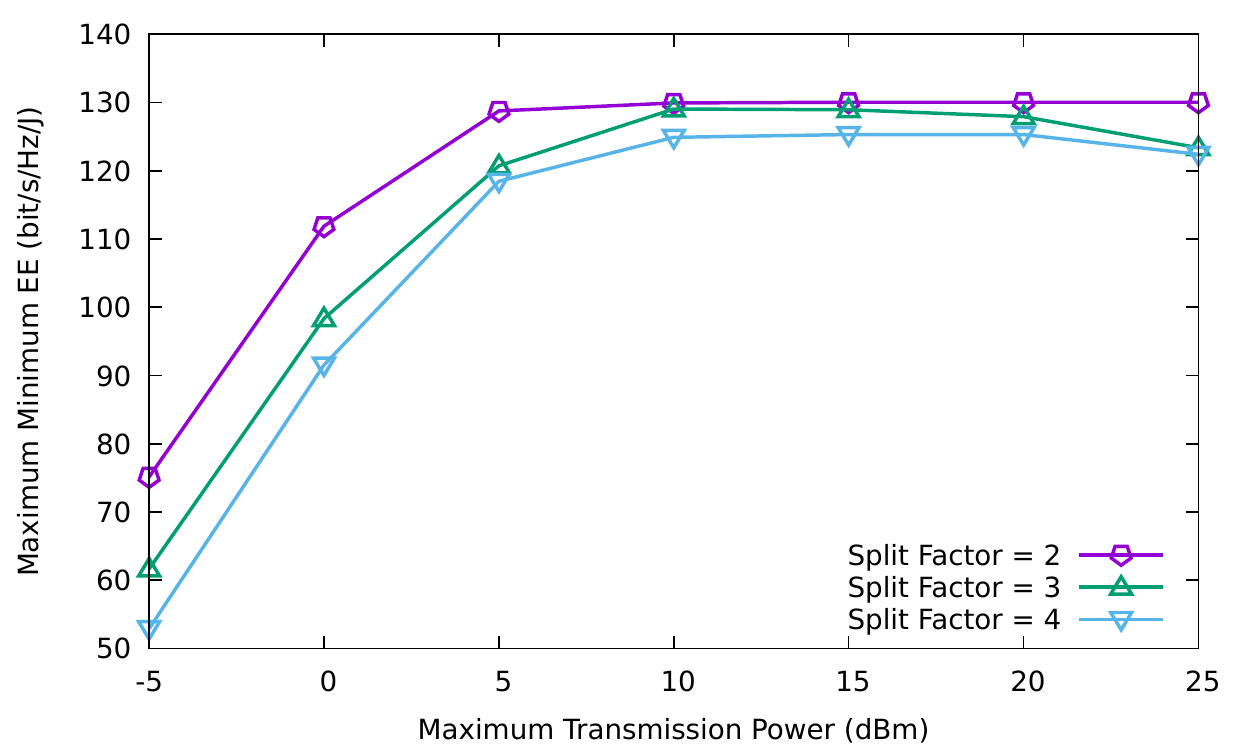}\hfil
	\includegraphics[width=9cm,height=5cm]{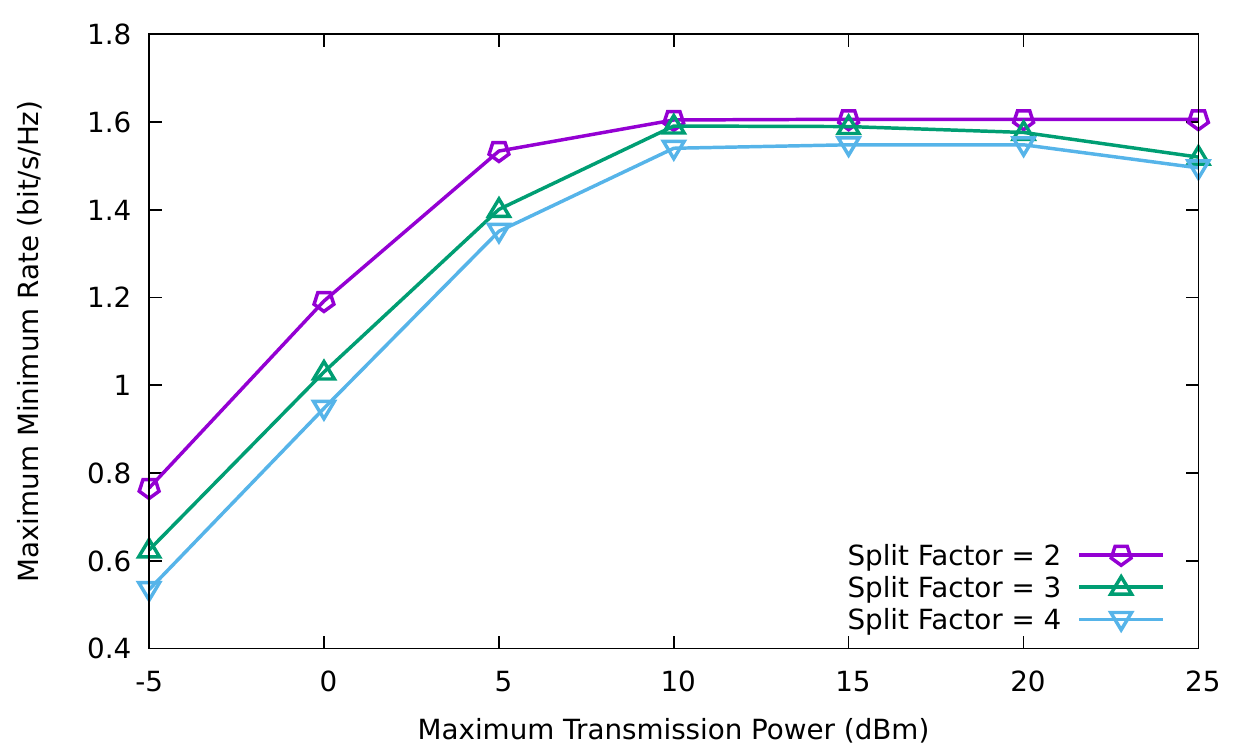}
	\caption{\label{fig:MTM-mee} MEE and Minimum rate vs. Transmission Power with Many-to-Many Matching.}
\end{figure*}

\subsubsection{Energy Efficiency and Rate Analysis} The sum-rate capacity and
the energy efficiency were analysed over the range of maximum allowed
transmission powers. Fig. 8 shows the aggregated rate for
different split factors. Clearly, the aggregate rate increases with the
maximum transmission power until the aggregate interference level prevents
further gains. Moreover, as D2MD is able to use more RBs, the aggregated rate
significantly improves. Notice that the behavior holds similar for GEE, as
Fig.8 illustrates, yet the effect of $s$ does not: as more
RBs are used, the needed power almost doubles from $122$ mW in DL for
$s = 2$ up to $242.4$ mW for $s = 4$, with a budget of
$-5$ dBm. Therefore, for the range of $[-5,10]\,$dBm we notice that
better GEE is achieved when $s = 2$. However, this changes for the range
between $15$ dBm and $25$ dBm, where higher rate values (saturation
points) and stable drawn power ($296.6$ --$ 298.6$ mW  for DL) are obtained.

\subsection{Discussion and Comparison Between Channel Allocation Scenarios}
Generally, and for all the presented algorithms, we notice that the clustering
techniques are not determinant to the fundamental performance for this type of
systems. In fact, the EE and rate curves show similar shapes and there are
only minor differences in their absolute values. In the one-to-one matching,
we assign a single RB per D2MD where $K$ is equal to $[5,10,15]$. Even though
the GEE decreases compared to the aggregated rate, yet the gain in term of
rate creates an acceptable balance, in particular, for low transmission power
budgets as illustrated in Fig. 4. The same conclusion
can be seen when applying the many-to-one matching algorithm. Here, with the
minimum number of RBs in the one-to-one case ($5$ RBs), the system supported
up to $25$ D2MD clusters due to the reuse factor. Fig. 6
shows the difference between the two algorithms. Obviously, the more D2MD
clusters are allocated to the same RB, the more transmission power will be
used by transmitters to satisfy the QoS constraints in the shadow of the
strong interference. Again, a small power budget creates a more controllable
interference even when the reuse factor $r$ is quiet high ($25$ D2MD cluster).
This can be seen in the Fig. 6, when maximum transmission
power is equal to $-5$ dBm, such as the difference between EE values when $r$
is equal to $1 ,2 ,3 ,4 ,5$ is growing as the budget do. In the final
algorithm, we introduce a split factor $s$ to investigate the effect of
allocating more resources to a D2MD cluster. Here, we clearly notice that the
more RBs, the more transmission budget is required to attain a high aggregated
rate and to create the required balance. In MEE, we aim to maximize the
minimum individual EE. Here, the user having the worst channel will always
consume more transmission power to satisfy the QoS constraints. This is
reflected in Fig. 7 with the one-to-one matching when
interference is from one side (a CUE or a cluster). Clearly, as $r$ grows, the
interference accumulates thus preventing the weakest user from achieving
higher rate. Differently, in Fig. 9  with the many-to-many
matching algorithm, the reuse factor is set to $2$ and the users are able to
distribute power and rate over $s$ RBs. This scenario provided a low
interference environment and high rate, mainly, for high power budgets.

As for the state information and signaling overhead required
by our algorithms, we should mention that for the computations involved in
the power control and channel assignment centralized problems, estimates of
the channel gains are needed. These can be acquired through pilot sequences
on a relatively slow time-scale in the case of slow fading. For the matching
theory based solution to the channel assignment problem, the necessary
channel state information is the aggregate interference level measured at
each receiver/group. Since a D2MD transmitter needs to learn its weakest
receiver (channel gain) and the aggregate interference level from its
co-channel transmitters, near or distant, the signaling overhead is
proportional to the number of receivers in its own group. 

\section{Conclusions}
\label{sec:conclusion}
In this paper, we modelled the joint power and resource allocation problem to
maximize the system and minimum individual energy efficiency.
We have shown that the behavior of both global and individual energy
efficiency is similar, with little dependence on the clustering of the
devices, provided that clusters have an area that is not comparable with the
cell area. We have provided a numerical framework for the planning and design
of D2D and D2MD communications in wireless networks. Results from this
framework indicate that fractional reuse of the resources, despite its
complexity, can be beneficial for increasing the global energy efficiency with
just minor degradation of the network sum-throughput. Therefore, a network
design based on the optimization of energy efficiency for a given target
throughput provides not only maximum energy efficiency, but also an adequate
control of the interference level suffered by the receivers. The combination
of matching theory and optimization used in the paper is essentially a
centralized planning solution. Future work includes investigating partially or
fully distributed implementations of the matching and optimization algorithms.
Another interesting research direction is the analysis of the trade-off
between optimal design and partial channel state information at the
transmitters as a way to reduce complexity. Finally, investigating the
performance for 5G wireless channels is another clear extension to this work.

\appendix[Complexity of the optimization problems]
\label{appendix}

\label{thm:np-hardness}
The RB sharing and power control problems stated in (10) and
 (11) are NP-hard.

The theorem is proved by showing that the RB sharing and power control problem
can be reduced in polynomial time to the integer partitioning problem: given a
set of positive integers $\{ n_1, \ldots, n_L \}$, determine if the set can be
partitioned in two subsets having the same sum. Thus, we can assume that
$\sum_{i = 1}^L n_i = 2 W$ for some integer $W$, since if the total sum were
odd the partition is clearly impossible. The proof works for the version that
maximizes the global energy efficiency, but is valid with straightforward
modifications to the other two versions, the maximization of the minimum EE,
and the maximization of the weighted sum of EEs.

\textbf{Proof}
 Recall that we have in our setting $M$ channels (or RBs) and $M$
CUEs. Assume for the rest of this proof that $K = 2$, i.e., only two groups
of D2D users, and that the reuse and split factors are $r = 2$, $s = M$. In
words, each channel will support only one CUE and one group, and each D2MD
group can potentially use all the channels. For simplicity in the notation,
but without loss of generality, we may also assume that each D2MD group
consist of only one pair of users (or even one user). This is not really a
restriction, since we could equivalently assume that all the point-to-point
channels between a given transmitter and the receivers in a group, say group
$j$, hace exactly the same quality.

Put $n_{\text{max}} = \max_{i = 1, \dots, L} n_i$, and set the noise power
to $\sigma^2 = 1/n_{\text{max}}$ and the SINR thresholds to
$\gamma_i = n_{\text{max}}$, for $i = 1, 2$. Now, set the channel gain
parameters in the following fashion (equations~\eqref{eq:sinr-d2d}
and~\eqref{eq:sinr-cue}). Fix $h_m = 1$ for all the CUEs $m = 1, \dots, M$,
and set also $h_{k,m} = 0$ for the D2MD transmitters $k = 1, 2$, so that the
D2MD groups do not cause interference to the CUEs. Set $h_{k,m,r} = 1$ for
$k = 1, 2$; $\beta_{k,m,r} = (n_m - 1) / n_{\text{max}}$; and finally,
$h_{j,m,r} = 1$ for the cross-talk channels between the clusters. Note that
the index $r$ in the latter parameters is void, since we are assuming a
single user per cluster, and also that $\beta_{k,m,r}$ is by definition a
number between zero and one.

Now, assume too that the consumption of the circuitry in the transmitters at
rest is zero, that the maximum transmission powers in a CU an a D2MD group
are $1$ and $W$, respectively, and that the rate of a CUE is equal to
$\log_2 (1 + n_{\text{max}})$. This value can only be reached if the CUEs
use their maximum transmission power, by the way. Suppose now that the two
groups use the same RB $i$ with powers $p_1$ and $p_2$. As the channel gains
for the cross-links are set to one, this means that
$p_1 \geq n_i + n_{\text{max}} p_2$, and
$p_2 \geq n_i + n_{\text{max}} p_1$.  Since all the $n_i \geq 1$, both
inequalities cannot hold simultaneously, and this means that the two groups
cannot share any RB and must be assigned to different channels. As a
consequence, since $h_{k,m} = 0$, the are not coupled by interference, and
the two groups can use up to the $M$ available channels in mutual exclusion,
i.e., with exactly one group (and one CUE) per RB\@. But if the SINR target f
group $1, 2$ has to be met, then its transmission power must be equal to
\begin{equation*}
n_{\text{max}} \left( \frac{1}{n_{\text{max}}} + \frac{n_i -
	1}{n_{\text{max}}} \right) = n_i.
\end{equation*}
This transmission power is limited to $W$ per channel, so a feasible
solution consisting in both groups using the $M$ channels, with only one
D2MD transmitter in each group, would use $2W$ at most. Since
$\sum_{i = 1}^M n_i = 2W$, a solution to the RB sharing and power control
problem would give us immediately a solution to the partitioning
problem. But the partitioning problem is
NP-complete~\cite{Garey2011}. Because the transformation of the original
problem into the partitioning problem is clearly polynomial, this proves
that the RB sharing and power control problem is NP-complete, and
consequently its optimization version is NP-hard.

\end{document}